\documentclass[aps,pre,groupedaddress,superscriptaddress,twocolumn]{revtex4-2}
\usepackage{amsmath}
\usepackage{float}
\usepackage{graphicx}
\usepackage{bm}
\usepackage{physics}
\usepackage{comment}
\usepackage{subcaption}
\usepackage[font=small,labelfont=bf]{caption}
\usepackage{hyperref}

\begin{document}

\title{Semiclassical analytical solutions of the eigenstate thermalization hypothesis in a quantum billiard}

\author{Yaoqi Ye}
\email{yyqnmcae@mail.ustc.edu.cn}
\affiliation{Department of Modern Physics, University of Science and Technology of China, Hefei 230026, China}
\author{Chengkai Lin}
\affiliation{Department of Modern Physics, University of Science and Technology of China, Hefei 230026, China}
\author{Xiao Wang}
\email{wx2398@mail.ustc.edu.cn}
\affiliation{Department of Modern Physics, University of Science and Technology of China, Hefei 230026, China}
\affiliation{
CAS Key Laboratory of Microscale Magnetic Resonance, University of Science and Technology of China, Hefei 230026, China}

\date{\today}

\begin{abstract}
We derive semiclassical analytical solutions for both the diagonal and off-diagonal functions in the eigenstate thermalization hypothesis (ETH) in a quarter-stadium quantum billiard. For a representative observable, we obtain an explicit expression and an asymptotic closed-form solution that naturally separate into a local contribution and a phase-space correlation term. These analytical results predict the band structure of the observable matrix, including its bandwidth and scaling behavior. We further demonstrate that our analytical formula is equivalent to the prediction of Berry’s conjecture. Supported by numerical evidence, we show that Berry’s conjecture captures the energetic long-wavelength behavior in the space of eigenstates, while our analytical solution describes the asymptotic behavior of the $f$ function in the semiclassical limit. Finally, by revealing the connection between the bandwidth scaling and the underlying classical dynamics, our results suggest that the ETH carries important physical implications in single-particle and few-body systems, where “thermalization” manifests as the loss of information about initial conditions.
\end{abstract}

\maketitle

\section{Introduction}
The eigenstate thermalization hypothesis (ETH) is a hypothesis about the properties of eigenstates in quantum chaotic systems.\cite{deutsch1991quantum,srednicki1994chaos,rigol2008thermalization} In recent years, it has attracted significant attention in condensed matter physics, statistical physics, high energy physics and quantum information~\cite{deutsch1991quantum,srednicki1994chaos,rigol2008thermalization,beugeling2014finite,d2016quantum,deutsch2018eigenstate,Mori_2018,RevModPhys.91.021001,PhysRevLett.130.140402,PhysRevLett.134.140404,bao2019eigenstate}.
 It is regarded as a crucial framework for understanding thermalization processes in isolated many-body quantum systems \cite{d2016quantum}. The ETH states that, in the energy eigenbasis $|E_i\rangle$, the matrix elements of a physical observable $O$ take the form \cite{srednicki1999approach}:
\begin{equation}
\langle E_i | O | E_j \rangle = O(E_i)\delta_{ij} + f(E_i, E_j) r_{ij},
\label{ETH}
\end{equation}
where $O(E_i)$ and $f(E_i, E_j)$ are smooth functions, and $r_{ij} = r_{ij}^*$ are random variables drawn from a Gaussian distribution with zero mean and unit variance.

Since its proposal, ETH has been extensively examined in both experiments and numerical simulations over the past three decades \cite{rigol2008thermalization, kaufman2016quantum, clos2016time, turner2018weak, jansen2019eigenstate, leblond2021universality, brenes2020eigenstate, mondaini2017eigenstate, leblond2019entanglement, richter2020eigenstate, deutsch2018eigenstate, rigol2012alternatives, de2015necessity, bernien2017probing, garrison2018does, khatami2013fluctuation, steinigeweg2013eigenstate}. More recently, analytical progress has been made in understanding the general properties of diagonal and off-diagonal elements \cite{mondaini2016eigenstate, dymarsky2022bound, wang2024semiclassicalstudydiagonaloffdiagonal,wang2025operatorweylsymbolapproacheigenstatethermalization}.

In this paper, we derive an explicit analytical expression for the off-diagonal ETH function of the observable $\hat{q}_x$ in a quantum billiard model, using a semiclassical approach based on recent analytical developments in ETH~\cite{wang2024semiclassicalstudydiagonaloffdiagonal, wang2025operatorweylsymbolapproacheigenstatethermalization}.
Our analytical solution for the off-diagonal function reveals further physical insights, as it can be decomposed into a local term and a phase-space correlation term.
An asymptotic closed-form expression of the $f$ function is also obtained, providing an explicit prediction for the bandwidth of the off-diagonal $f$ function.
With these previously unavailable analytical results, we identify scaling behaviors for the bandwidth of the off-diagonal function that are consistent with the intuition from the corresponding classical dynamics.
Furthermore, we demonstrate that our analytical prediction is equivalent to Berry’s conjecture~\cite{berry1977regular}.
Finally, supported by numerical evidence, we argue that Berry’s conjecture describes the energetic long-wavelength behavior in the space of eigenstates, implying that our analytical solution for the off-diagonal function captures the correct behavior in the semiclassical limit $\hbar \to 0$.

We provide the first analytical solutions for off-diagonal elements of an observable since ETH was proposed. We also provide semiclassical prediction and numerical simulation for diagonal elements. Furthermore, our results demonstrate that features of ETH are not restricted to many-body systems but also emerges in single-particle chaotic systems with fewer degrees of freedom.

\section{Model}

\begin{figure}[H]
\centering
\includegraphics[width=0.3\textwidth]{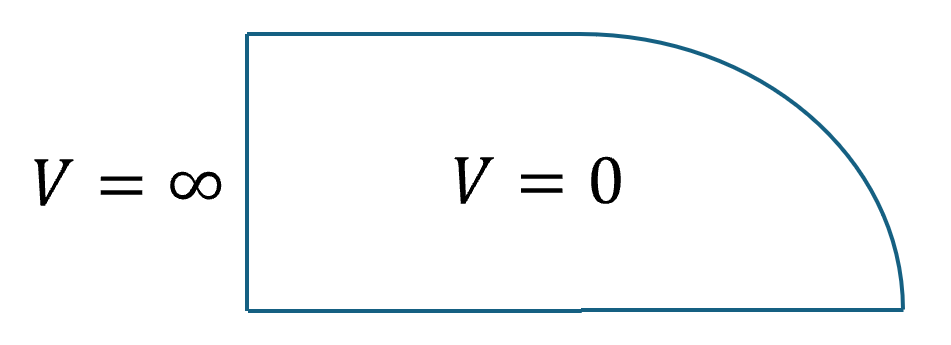}
\caption{Geometry of the quarter-stadium billiard with width $l=2$ and height $h=1$.}
\label{model}
\end{figure}

We consider the quantum billiard defined on a quarter-stadium geometry. This model describes the motion of a particle confined in a two-dimensional infinite potential well shaped as a quarter stadium (see Fig.~\ref{model}).

The system is classically chaotic \cite{bohigas1984characterization}, and its quantum counterpart is expected to follow Berry's random-wave conjecture \cite{berry1977regular,berry1989quantum}.

For an eigenstate with energy $E_i$, where $i$ denotes the level index counted from the ground state, one has 
\begin{equation}
    E_i \propto \hbar^2 m^{-1} S_\Omega^{-1},
    \label{E_i}
\end{equation}
where $\Omega$ denotes the quarter-stadium region and $S_\Omega$ is its area. Accordingly, the density of states scales as $\rho_{\text{dos}} \propto \hbar^2 m^{-1} S_\Omega^{-1}$. (see Appendix~\ref{scaling})

\section{Semiclassical expression for ETH in the quantum billiard model}

We use the following exact expression for the off-diagonal matrix elements as derived in Ref.~\cite{wang2025operatorweylsymbolapproacheigenstatethermalization}:
\begin{equation}
\begin{split}
&| \langle E_i | O | E_j \rangle |^2 = \int d\bm{p} \, d\bm{q} \int d\bm{p}' \, d\bm{q}' \, \\
&\quad \times J \left( \bm{p}, \bm{q}, \frac{\bm{q}'}{\hbar}, \frac{\bm{p}'}{\hbar} \right) 
   W_i \left( \bm{p} - \frac{\bm{p}'}{2}, \bm{q} - \frac{\bm{q}'}{2} \right) \\
&\quad \times W_j \left( \bm{p} + \frac{\bm{p}'}{2}, \bm{q} + \frac{\bm{q}'}{2} \right),
\end{split}
\label{Osq}
\end{equation}
with \( i \neq j \), where
\begin{equation}
\begin{split}
&J \left( \bm{p}, \bm{q}, \bm{k}_p, \bm{k}_q \right) = \frac{1}{(2\pi\hbar)^f} \int d\bm{\tilde{p}} \, d\bm{\tilde{q}} \, \\
&\quad \times e^{-i \bm{k}_p \cdot \bm{\tilde{p}}} e^{-i \bm{k}_q \cdot \bm{\tilde{q}}} 
   O_{\mathrm{\scriptscriptstyle W}} \left( \bm{p} + \tfrac{1}{2} \bm{\tilde{p}}, \bm{q} - \tfrac{1}{2} \bm{\tilde{q}} \right) \\
&\quad \times O_{\mathrm{\scriptscriptstyle W}} \left( \bm{p} - \tfrac{1}{2} \bm{\tilde{p}}, \bm{q} + \tfrac{1}{2} \bm{\tilde{q}} \right).
\end{split}
\label{J}
\end{equation}

Here, $O_{\mathrm{\scriptscriptstyle W}}(\bm{p}, \bm{q})$ is the \textit{Weyl symbol}~\cite{weyl1927quantenmechanik} of the operator $O(\hat{\mathbf{p}}, \hat{\mathbf{q}})$, defined by
\begin{equation}
O_{\mathrm{\scriptscriptstyle W}}(\bm{p}, \bm{q})
= \int d\bm{r} \, e^{-i \bm{p} \cdot \bm{r} / \hbar}
\left\langle \bm{q} + \frac{\bm{r}}{2} \middle| 
O(\hat{\bm{p}}, \hat{\bm{q}}) 
\middle| \bm{q} - \frac{\bm{r}}{2} \right\rangle .
\label{eq:WeylSymbol}
\end{equation}
The explicit expression of $O_{\mathrm{\scriptscriptstyle W}}(\bm{p}, \bm{q})$ is obtainable by writing 
the operator function $O(\hat{\bm{p}}, \hat{\bm{q}})$ as a sum of Weyl-ordered 
functions~\cite{wang2024semiclassicalstudydiagonaloffdiagonal}. In our case, where we choose the observable $\hat{q}_x$, $O_{\mathrm{\scriptscriptstyle W}}(\bm{p}, \bm{q})$ simply equals to $q_x$.

Eq.~\eqref{Osq} explicitly demonstrates that the $f$ function is determined by the two point correlation of the wavefunction, since Wigner function is its Fourier transformation. It is generally assumed that for chaotic systems the averaged Wigner function over a narrow energy window semiclassically corresponds to the energy surface in phase space \cite{berry1977regular, voros1976semi,voros1977asymptotic} defined by $ H_{cl}(\bm{p}, \bm{q}) = E_i $, namely,
\begin{equation}
\overline{W}_i(\bm{p}, \bm{q}) = \frac{\delta\big(H_{cl}(\bm{p}, \bm{q}) - E_i\big)}{S(E_i)},
\label{Wsc}
\end{equation}
where \( S(E_i) \) is the normalization factor, denoting the volume of the energy shell in phase space (Eq.~\ref{Wsc} exhibits a divergence in the second-order normalization; see Appendix~\ref{2nd order normalization} for details.):
\begin{equation}
S(E_i) = \int d\bm{p} \, d\bm{q} \, \delta\big(H_{cl}(\bm{p}, \bm{q}) - E_i\big).
\label{S}
\end{equation}
Note that Eq.~\eqref{Wsc} has direct connection with Berry's conjecture (see Appendix~\ref{Berry and Wigner}). Since Wigner function has been averaged over a narrow energy shell, the off-diagonal elements computed using Eq.~\eqref{Wsc} can be regarded as good approximations of the $f$ function. Moreover, this approximation is expected to become more accurate at higher energy levels, where the semiclassical theory and Berry's conjecture are better satisfied. In the following contexts, we will simply call Eq.~\eqref{Wsc} Wigner function, denoted as $W_i$. $S(E_i)=2\pi mS_\Omega$ (see Appendix~\ref{derivation of S}) is a constant in our model.

\section{Analytical solution for off-diagonal elements of $\hat{q}_x$}
\label{analytical analysis}
The observable under consideration is the x-component of real space, namely,
\begin{equation}
O_{\mathrm{\scriptscriptstyle W}}(\bm{p},\bm{q})=q_x
\label{Ox}
\end{equation}

Substituting Eq.~\eqref{Ox} and Eq.~\eqref{Wsc} into Eqs.~\eqref{J}\eqref{Osq}, we get (detailed derivations can be found in Appendix~\ref{derivation of Eq9})
\begin{equation}
\begin{split}
&\overline{|\bra{E_i}\hat{q}_x\ket{E_j}|^2}=\frac{1}{(1+\pi/4)^2}\int d\bm{z} z_x^2 \\
&\times\int d\bm{\widetilde{z}} \,J_0(a|\bm{\widetilde{z}}|\sqrt{2mE_i}/\hbar)J_0(a|\bm{\widetilde{z}}|\sqrt{2mE_j}/\hbar)\\
&-\frac{1}{4(1+\pi/4)^2}\int d\bm{z} d\bm{\widetilde{z}}\,  \bm\widetilde{z}_x^2 J_0(a|\bm{\widetilde{z}}|\sqrt{2mE_i}/\hbar)\\
&\times J_0(a|\bm{\widetilde{z}}|\sqrt{2mE_j}/\hbar)\\
\end{split}
\label{qx_sq}
\end{equation}
Here we rescale the variables $\tilde{\bm{z}}=\frac{\tilde{\bm{q}}}{a}$ and $\bm{z}=\frac{\bm{q}}{a}$ in Eq.~\eqref{qx_sq_app} to remove their dependence on the billiard size, where $a=\sqrt{\frac{S_\Omega}{1+\pi/4}}$ , with the denominator corresponding to the area of the geometry shown in Fig.~\ref{model}. $|\tilde{\bm{z}}|$ is the magnitude of $\bm{\tilde{z}}$, and we employ polar coordinates for $\bm{\tilde{z}}$.

The two terms in Eq.~\eqref{qx_sq} originate from the product of two Weyl symbols in Eq.~\eqref{J}, which is equal to $q_x^2-\frac{\tilde{q}_x^2}{4}$. The second term stands for the correlation of the quantum observable in phase space, while the first term corresponds to the localized part in phase space. In our numerical results (Fig.~\ref{fig:offdiag}, we also find that the local term is the leading term in Eq.~\eqref{qx_sq}, greater than the correlation term by about one order of magnitude.

Eq.~\eqref{qx_sq} predicts that the observable matrix $\bra{E_i}\hat{q}_x\ket{E_j}$ exhibits a band structure (see Fig.~\ref{fig:offdiag}), with dominant components concentrated near the diagonal, where $E_i \approx E_j \approx \overline{E} = (E_i + E_j)/2$. Mathematically, this arises because, as the energy difference $|E_i - E_j|$ increases, the two Bessel functions in the integrand oscillate with increasingly different frequencies, resulting in a reduced contribution to the integral. By expressing the argument of the Bessel function as $\frac{a|\tilde{\bm{z}}|}{\hbar}\sqrt{\frac{2m}{E_{i,j}}}E_{i,j}$, one can see that the off-diagonal function $f(E_i, E_j) \approx \overline{|\bra{E_i}\hat{q}_x\ket{E_j}|^2}$ scales as $\hbar \sqrt{\frac{\overline{E}}{2mS_\Omega}}$ in both the $E_i$ and $E_j$ directions when $E_i\approx E_j$. This implies that the bandwidth is also proportional to this quantity. Within the billiard region, $p = \sqrt{2mE} = \frac{1}{m} |\nabla H_{\mathrm{cl}}|$, where our prediction for the relation between the bandwidth and $|\nabla H_{\mathrm{cl}}|$ agrees with the argument presented in Ref.~\cite{wang2025operatorweylsymbolapproacheigenstatethermalization}.

It is generally expected that the thermalization time scales as  $\tau\sim\hbar/w_b$\cite{wang2022eigenstate,schiulaz2019thouless,serbyn2017thouless,vsuntajs2020quantum,kos2021chaos,wang2025operatorweylsymbolapproacheigenstatethermalization}, where $w_b$ is the bandwidth. In many-body systems, thermalization is often characterized by the loss of memory of the initial condition, a concept that also applies to our single-particle case. Our analytical solution predicts the scaling behavior of thermalization time $\tau \sim (\frac{p}{m})^{-1}\sqrt{S_\Omega}=v^{-1}\sqrt{S_\Omega}$, where $v$ denotes the particle velocity, consistent with the classical intuition that a billiard particle moving faster or confined within a smaller cavity—thus experiencing more frequent collisions with the walls—loses memory of its initial condition more rapidly.

\section{Asymptotical Analytical Form}
We now choose the billiard size to be the same in Fig.~\ref{model} where $S_\Omega=1+\pi/4$ and $a=1$, and introduce a cutoff $Q$ for the integration domain of $|\bm{\tilde{q}}'|$, where $Q$ is chosen to be of the same order as the linear size of the quarter-stadium billiard. Considering high energy levels, Eq.~\eqref{qx_sq} becomes (detailed derivations can be found in Appendix~\ref{derivation of Eq10})
\begin{widetext}
\begin{equation}
\begin{split}
&\overline{|\bra{E_i}\hat{q}_x\ket{E_j}|^2}\approx\frac{\hbar^2}{mS_\Omega(E_iE_j)^{\frac{1}{4}}}\biggl\{\left[\frac{1}{S_\Omega}\left(\frac{5\pi}{16}+1\right)-\frac{1}{4}Q^2\right]\frac{\sin\left[\frac{\sqrt{2m}}{\hbar}(\sqrt{E_i}-\sqrt{E_j})Q\right]}{\sqrt{E_i}-\sqrt{E_j}}\\
&-\frac{1}{2}\frac{\hbar}{\sqrt{2m}}Q\frac{\cos\left[\frac{\sqrt{2m}}{\hbar}(\sqrt{E_i}-\sqrt{E_j})Q\right]}{(\sqrt{E_i}-\sqrt{E_j})^2}+\frac{1}{2}\left(\frac{\hbar}{\sqrt{2m}}\right)^2\frac{\sin\left[\frac{\sqrt{2m}}{\hbar}(\sqrt{E_i}-\sqrt{E_j})Q\right]}{(\sqrt{E_i}-\sqrt{E_j})^3}\biggr\}.
\end{split}
\label{qx_sq_analy}
\end{equation}
\end{widetext}

\begin{figure}[htbp]
    \centering
    \includegraphics[width=0.45\textwidth]{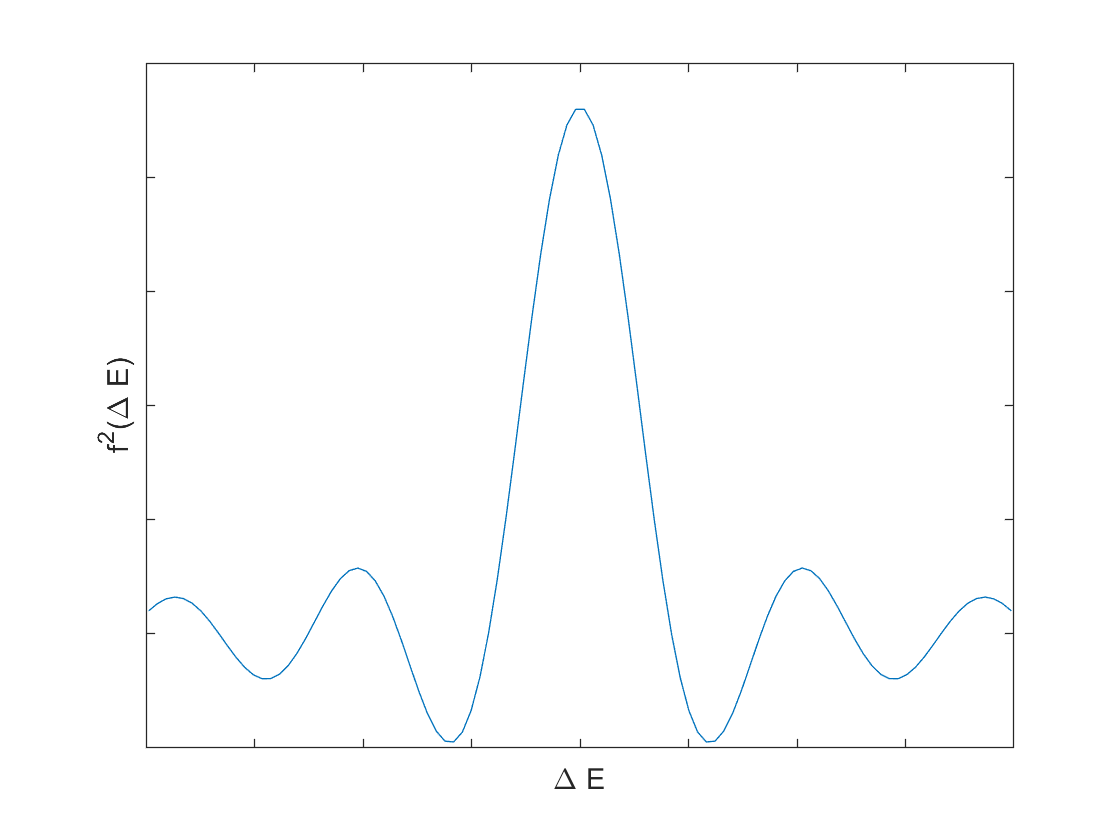}
    \caption{Typical behavior of Eq.~\eqref{qx_sq_analy} as a function of $\Delta E$.}
    \label{fig:typi_as}
\end{figure}
It is straightforward to see that the magnitude of lowest order $\mathcal{O}(\hbar^2)$ term is proportional to $\rho_{dos}^{-1}$. The width of the main peak in the middle, with typical behavior shown in Fig.~\ref{fig:typi_as}, can be inferred from the period of those trigonometric functions. The expression inside these trigonometric functions can be written as $\frac{\sqrt{2m}Q}{\hbar(\sqrt{E_i}+\sqrt{E_j})}\Delta E$, where $\Delta{E}=E_i-E_j$. Thus, the period is $\frac{2\pi\hbar}{\sqrt{2m}Q}(\sqrt{E_i}+\sqrt{E_j})\approx\frac{2\pi\hbar}{Q}\sqrt{\frac{2E}{m}}$, where $E=(E_i+E_j)/2$. The bandwidth approximately equals to half the period  $\frac{\pi\hbar}{Q}\sqrt{\frac{2E}{m}}\approx \pi\hbar\sqrt{\frac{2E}{mS_\Omega}}$, where we choose $Q=\sqrt{S_\Omega}$. Note that the scaling behavior of the bandwidth is consistent with the analysis presented in Sec.~\ref{analytical analysis}.

\section{Analytical Solution for Diagonal Elements of $q_x$}
Using the semiclassical expression of the diagonal elements in Ref.~\cite{wang2024semiclassicalstudydiagonaloffdiagonal}
\begin{equation}
O(E_i) \simeq \frac{1}{S(E_i)} \int d\bm p \, d\bm q \, \delta \!\big( H_{\mathrm{cl}}(\bm p,\bm q) - E_i \big) O_{\mathrm{\scriptscriptstyle W}}(\bm p,\bm q),
\label{O_diag}
\end{equation}
one can compute the diagonal function for $\hat{q}_x$ directly.

The result for diagonal elements is a constant 
\begin{equation}
\overline{|\bra{E_i}\hat{q}_x\ket{E_i}|}=\frac{10+3\pi}{12+3\pi}\approx0.967.
\label{diag}
\end{equation}

\section{Numerical simulation}
\label{section_NS}
\begin{figure}[htbp]
\includegraphics[trim=0 200 0 200,clip,width=0.45\textwidth]{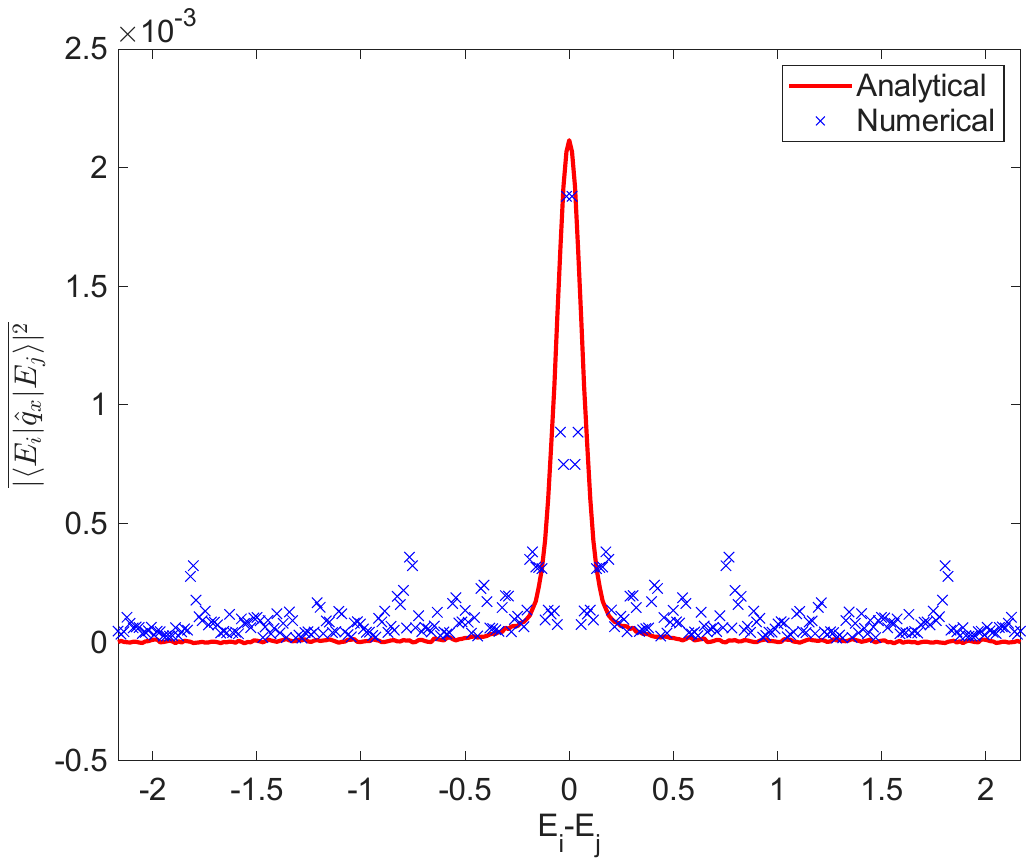}
\caption{The red line represents our analytical result from Eq.~\eqref{qx_sq}, while the blue dots show the numerical results for the squared off-diagonal 
$f$ function.}
\label{fig:offdiag}
\end{figure}

\begin{figure}[htbp]
    \includegraphics[trim=0 200 0 200,clip,width=0.45\textwidth]{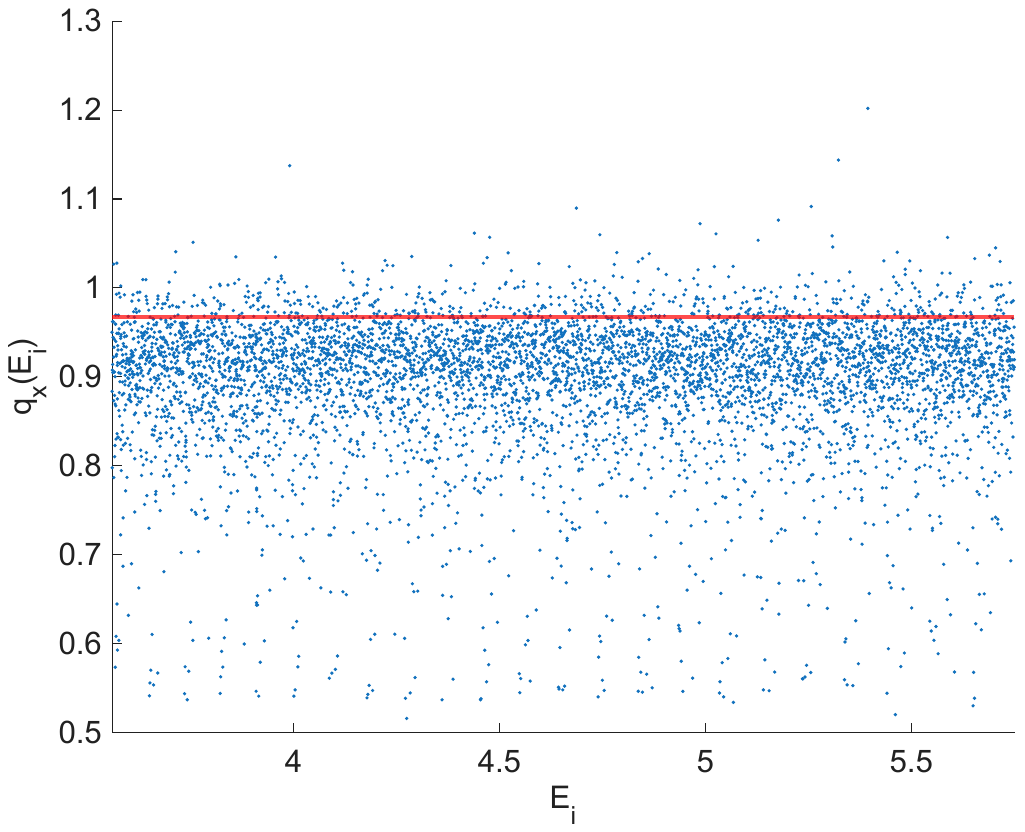}
    \caption{The blue dots are the numerical results for diagonal elements $\bra{E_i}\hat{q}_x\ket{E_i}$}. The red line is our theoretical prediction with averaged Wigner function.
    \label{fig:diag}
    
\end{figure}

We numerically study ETH in the quarter stadium billiard model using the scaling method \cite{vergini1995calculation}. In our simulation, we set $\hbar = 0.01$. We considered the energy range $3.5610 \sim 5.7469$, while energy levels are picked sparsely with equal space to reduce computational complexity. The numerical results for the off-diagonal $f$ function of $\hat{q}_x$ are shown in Fig.~\ref{fig:offdiag} together with our analytical prediction, where we restrict the integration domain for the observable to be within the billiard, i.e. $\bm{q}\pm\frac{\bm{\tilde{q}}}{2}\in \Omega$. Note that the restriction $\bm{q}\in\Omega$, imposed by the infinite potential wall in the Hamiltonian, is automatically fulfilled by the preceding condition. Numerical simulation for diagonal elements is also shown in Fig.~\ref{fig:diag}

Our analytical solution captures the main behavior of the off-diagonal $f$ function and reproduces the peak magnitude observed in the numerical results. The asymptotical analytical form predicts the bandwidth to be about $0.07$. Those deviations of the numerical results from our analytical prediction reveal fluctuations and correlations among eigenstates, which are neglected by our approximation of the Wigner function. In Appendix~\ref{Berry and Wigner}, we prove the equivalence of our analytical prediction and Berry's conjecture. It turns out that Berry's conjecture stands for the limit of long-wavelength behavior in the space of eigenstates. If one average over more eigenstates when computing the $f$ function, the numerical result will match better with our analytical solution, while also being more computationally expensive. This is equivalent to averaging over a small energy shell in the semiclassical limit $\hbar\to0$. We will demonstrate this in the next section, by providing numerical evidence.

\section{Berry's Conjecture as the Energetic Long-Wavelength Behavior}
\begin{figure}[htbp]
    \centering
    \begin{subfigure}[b]{0.23\textwidth}
        \centering
        \includegraphics[width=\textwidth]{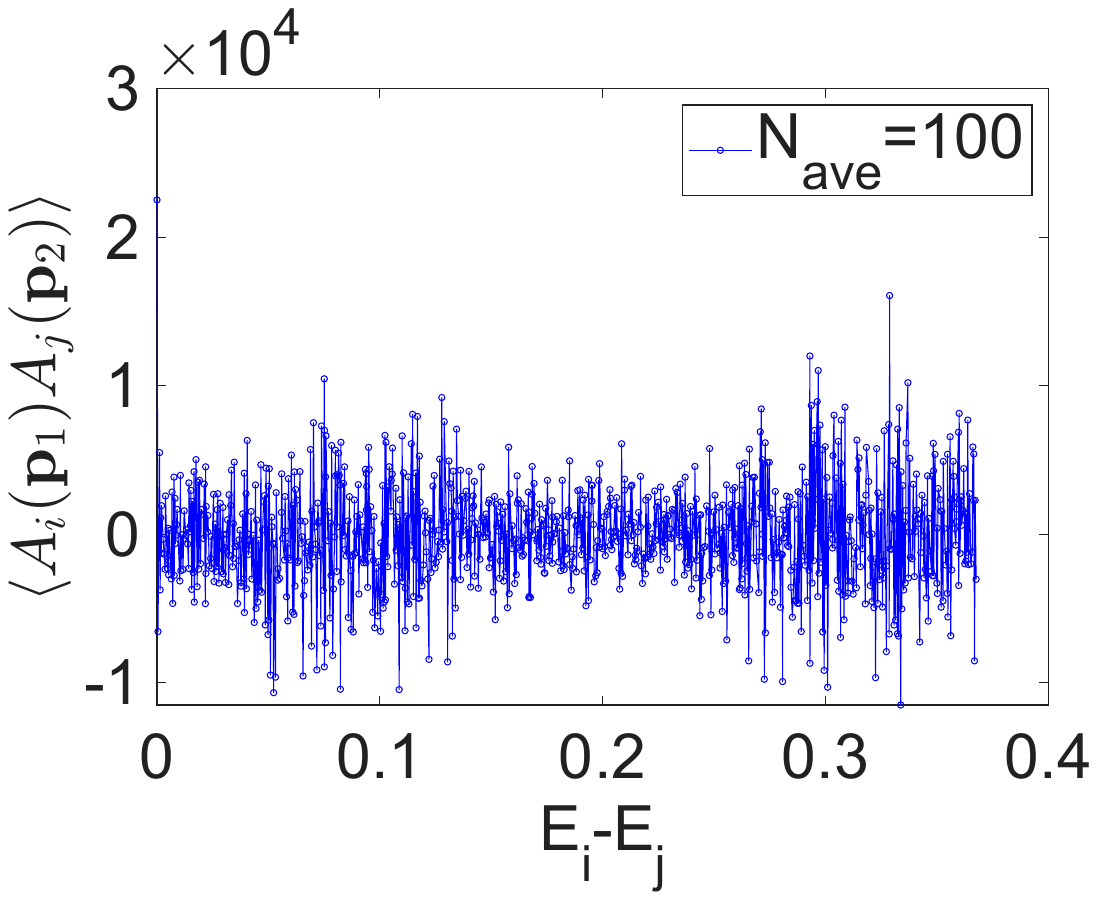}
        \label{fig:1a}
    \end{subfigure}
    \hfill
    \begin{subfigure}[b]{0.23\textwidth}
        \centering
        \includegraphics[width=\textwidth]{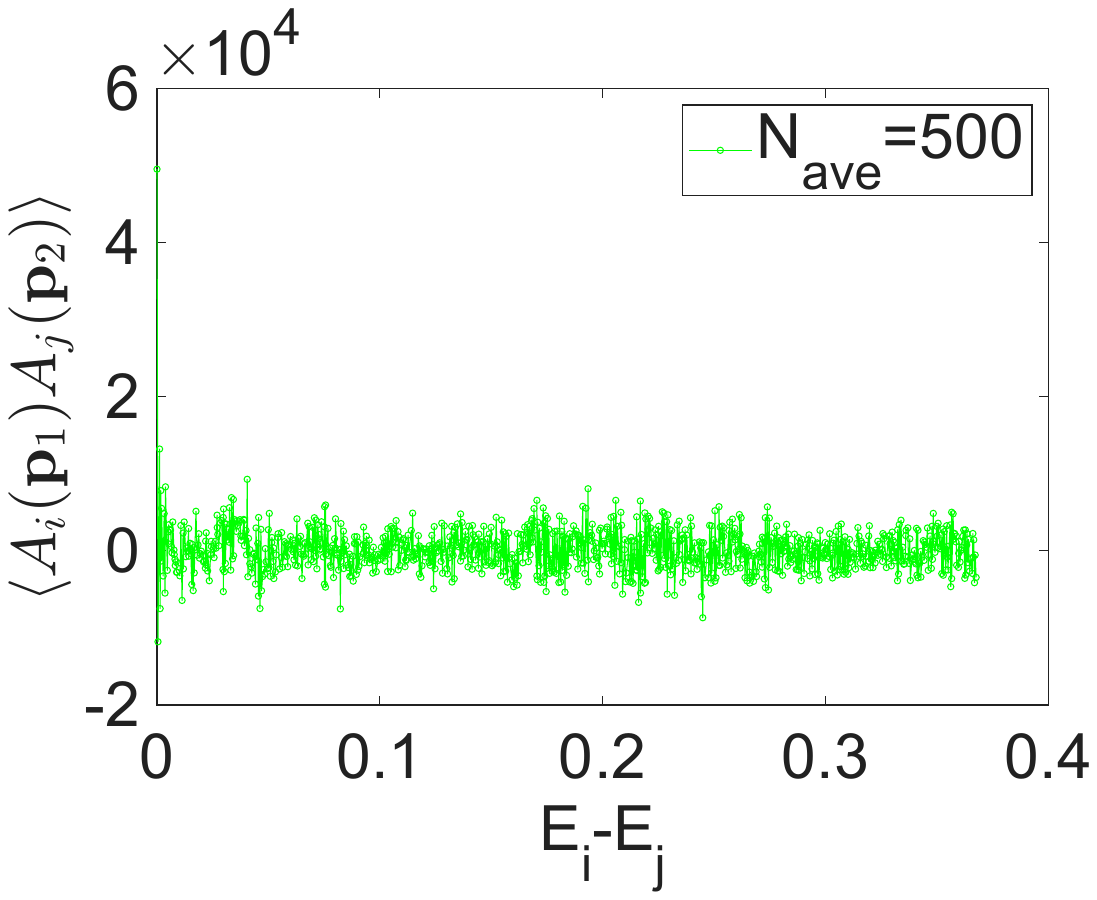}
        \label{fig:1b}
    \end{subfigure}

    \vspace{-10em}
    \begin{subfigure}[b]{0.23\textwidth}
        \centering
        \includegraphics[width=\textwidth]{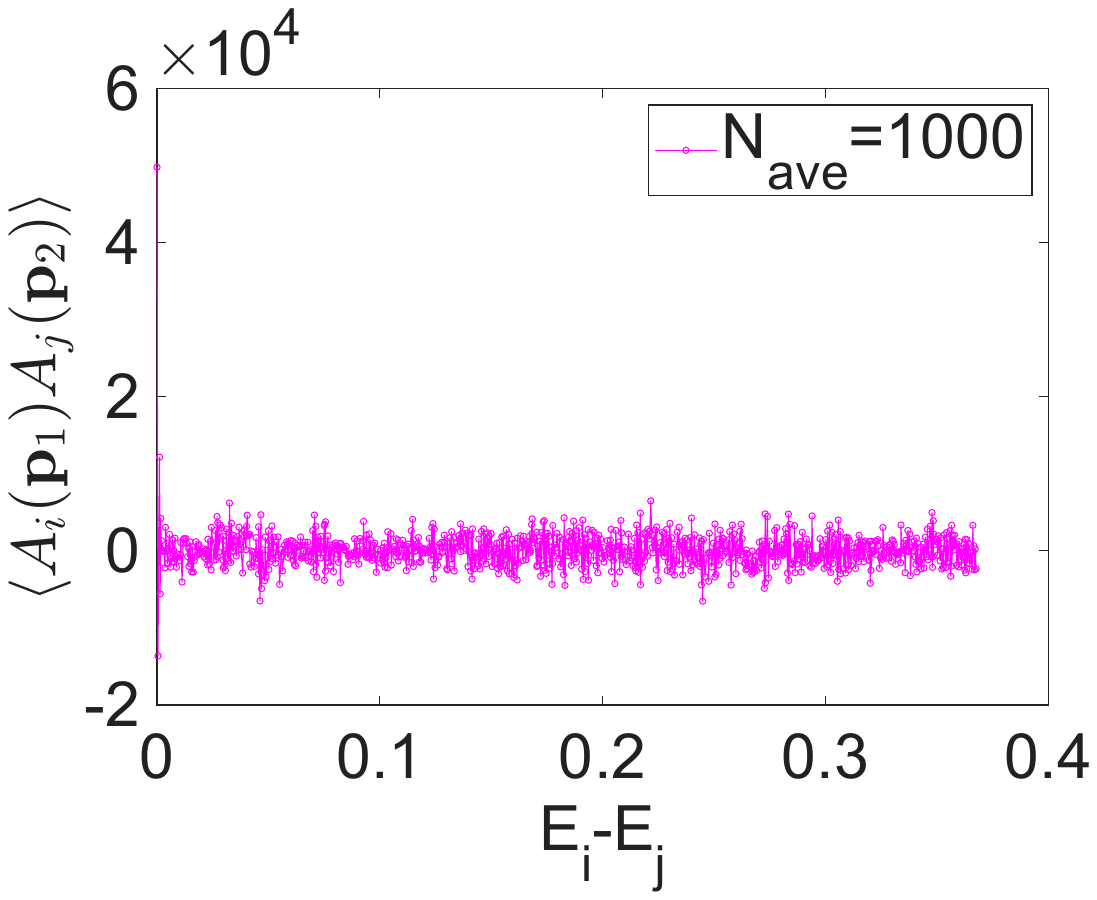}
        \label{fig:1c}
    \end{subfigure}
    \hfill
    \begin{subfigure}[b]{0.23\textwidth}
        \centering
        \includegraphics[width=\textwidth]{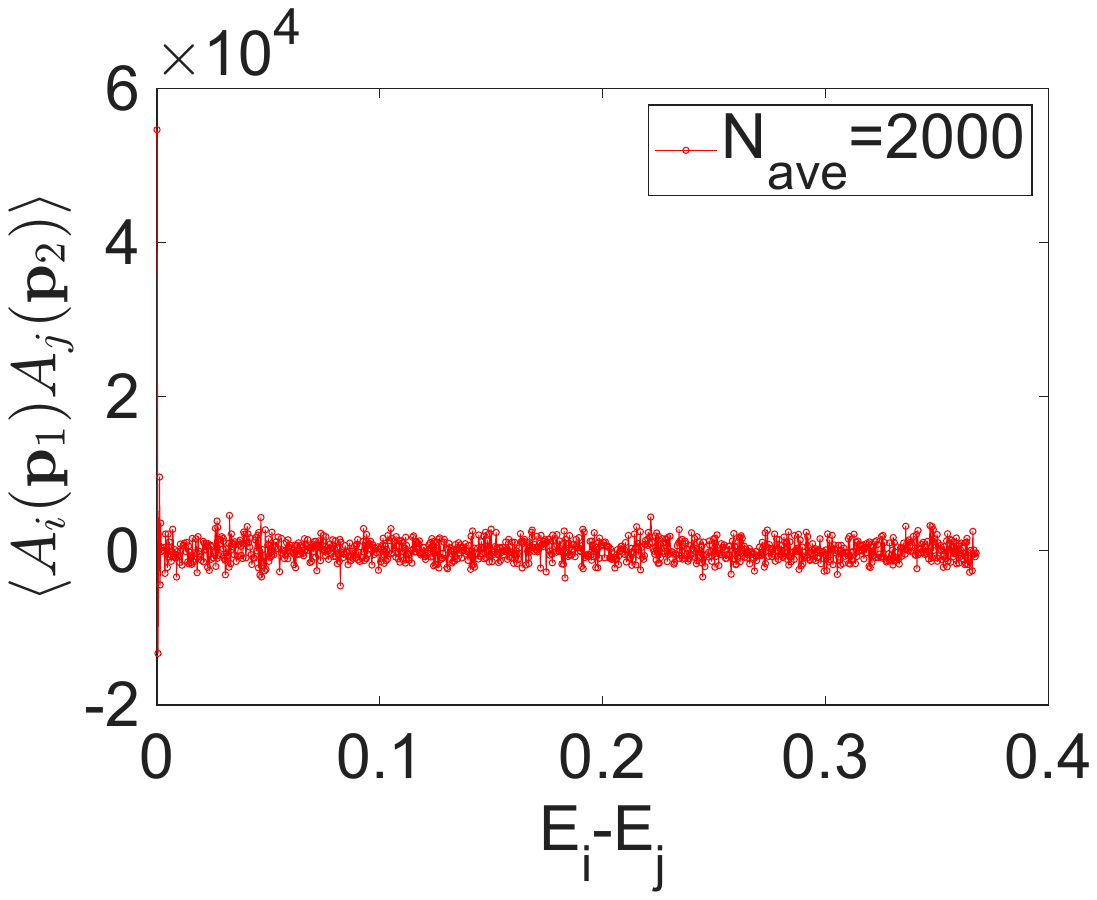}
        \label{fig:1d}
    \end{subfigure}
    \vspace{-5em}
    \caption{Here is the result for $\langle A_i(\bm{p}_1)A_j(\bm{p}_2)\rangle$ while setting $\bm{p}_1=\bm{p}_2$. The average is taken over an energy shell consisting $N_{ave}$ energy levels.}
    \label{fig:AAE}
\end{figure}

The numerical scaling method solves the Schrödinger equation of the billiard in momentum space, enabling us to directly look into Berry's conjecture, which assumes there to be no correlation among or within the eigenstates,i.e.\cite{srednicki1994chaos}
\begin{equation}
    \langle \mathcal{A}_i(\bm{p}_1)\mathcal{A}_j(\bm{p}_2)\rangle= \frac{\delta_{ij}\delta(\bm{p}_1+\bm{p}_2)}{\delta(\bm{p}_1^2-\bm{p}_2^2)},
    \label{Berry}
\end{equation}
where $\mathcal{A}_i^*(\bm{p})=\mathcal{A}_i(-\bm{p})=\mathcal{A}_i(\bm{p})$ in our model, and the average is taken over a small energy shell.
$A_i(\bm{p})=\mathcal{A}_i(\bm{p})\delta(\bm{p}^2-2mE_i)$ is the complex amplitude on $\ket{\bm{p}}$ in $i^{th}$ eigenstate, which is a real number in quarter stadium billiard, where the eigenstate is 
\begin{equation}
    \ket{E_i}=\mathcal{N}_i\int d\bm{p}\,\mathcal{A}_i(\bm{p})\delta(\bm{p}^2-2mE_i)\ket{\bm{p}}.
    \label{EigenBerry}
\end{equation}
$\mathcal{N}_i$ is the normalization factor.
Eq.~\eqref{Berry} can directly leads to Eq.~\eqref{qx_sq}(see detailed derivations in Appendix~\ref{Berry and Wigner}). In Fig.~\ref{fig:AA} and \ref{fig:AAE}, we numerically compute $\langle A_i(\bm{p}_1)A_j(\bm{p}_2)\rangle$ in the cases where $i=j$ and $\bm{p}_1=\bm{p}_2$.

\begin{figure}[htbp]
    \centering
    \begin{subfigure}[b]{0.23\textwidth}
        \centering
        \includegraphics[width=\textwidth]{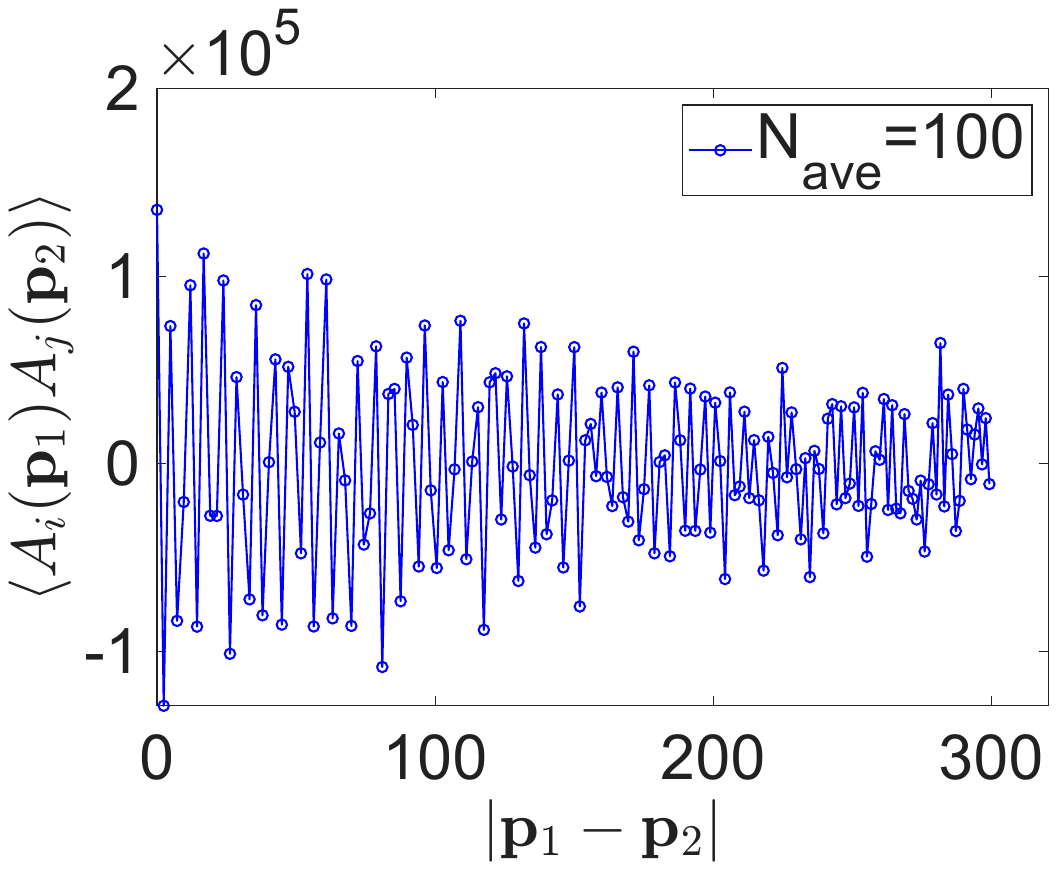}
        \label{fig:1a}
    \end{subfigure}
    \hfill
    \begin{subfigure}[b]{0.23\textwidth}
        \centering
        \includegraphics[width=\textwidth]{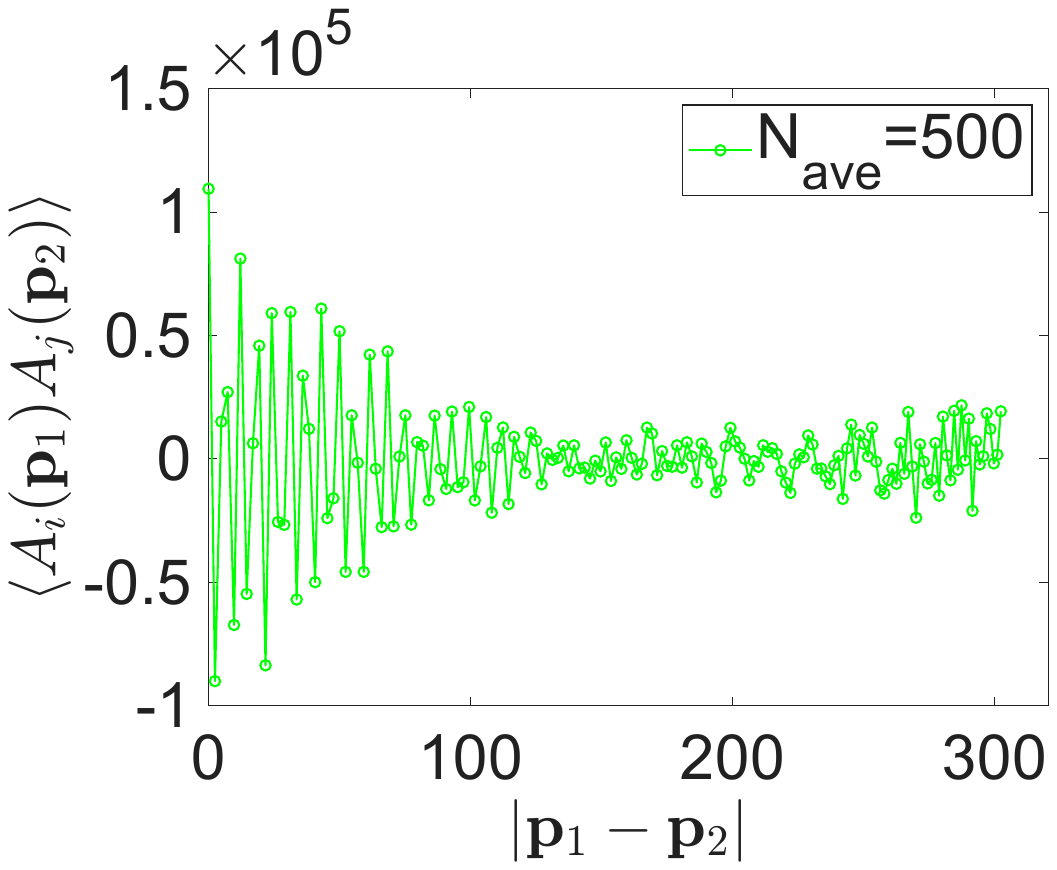}
        \label{fig:1b}
    \end{subfigure}
    \vspace{-10em}

    \begin{subfigure}[b]{0.23\textwidth}
        \centering
        \includegraphics[width=\textwidth]{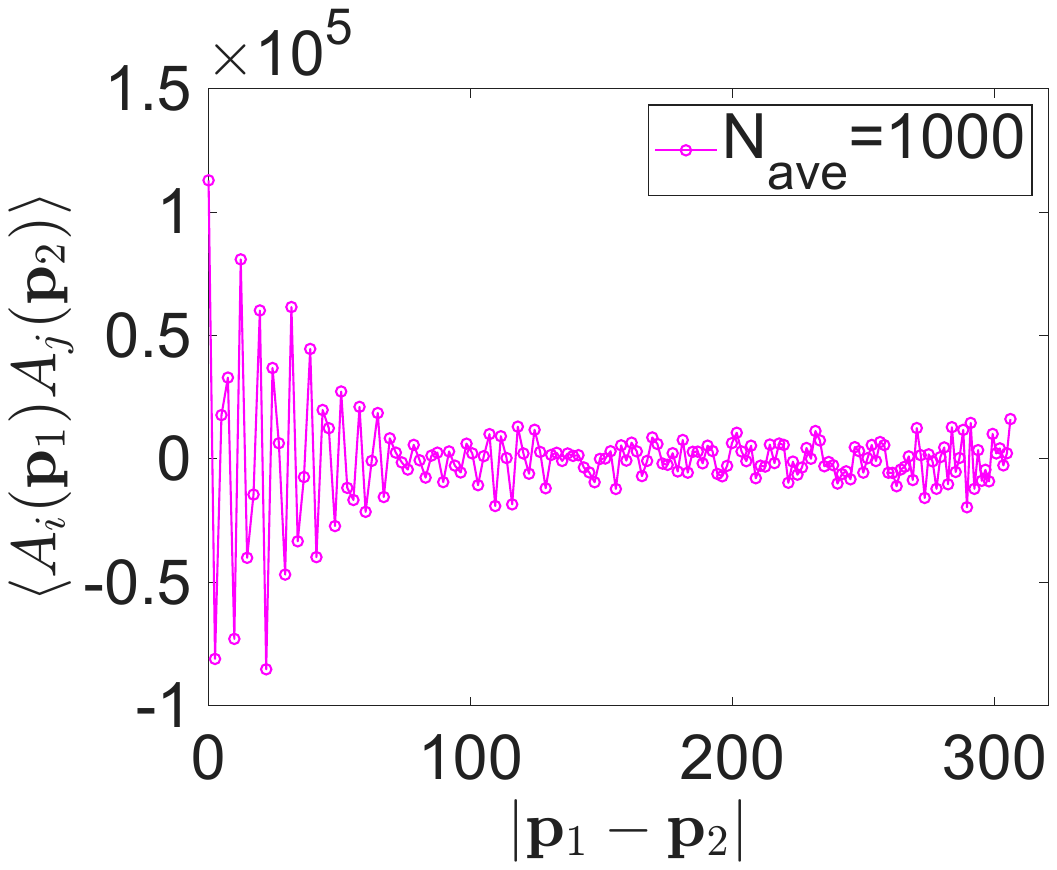}
        \label{fig:1c}
    \end{subfigure}
    \hfill
    \begin{subfigure}[b]{0.23\textwidth}
        \centering
        \includegraphics[width=\textwidth]{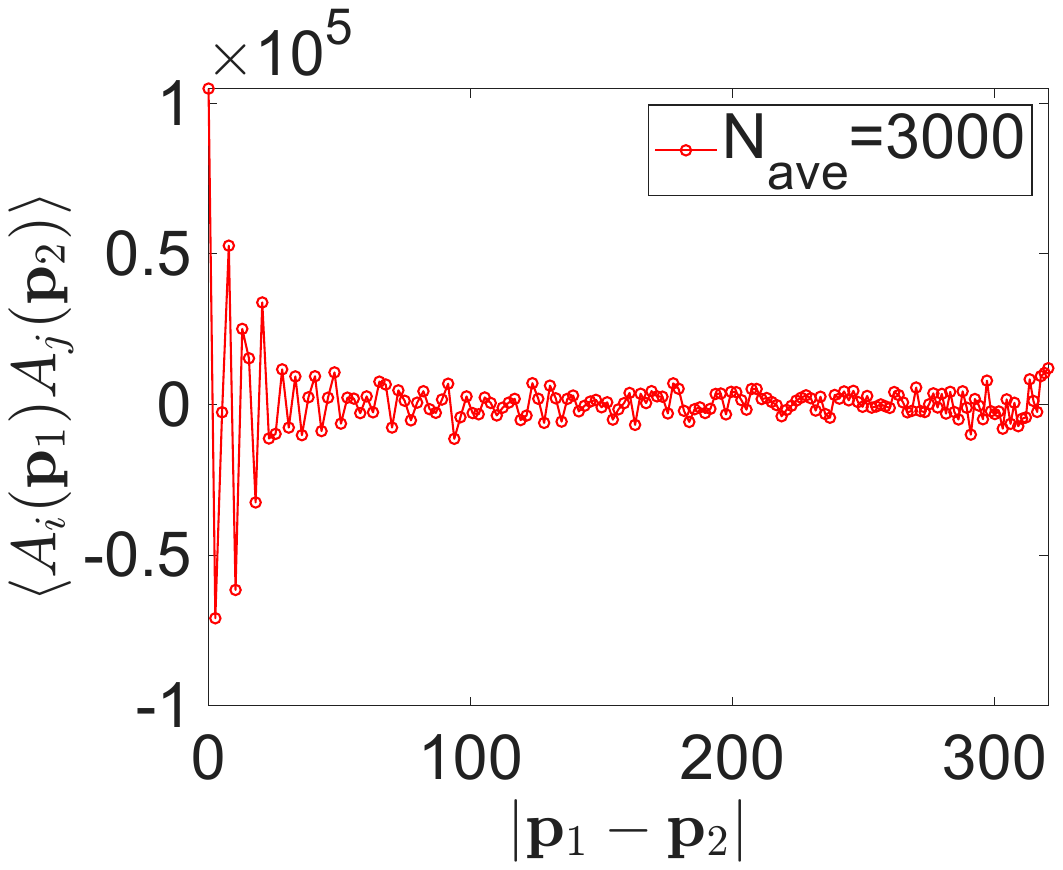}
        \label{fig:1d}
    \end{subfigure}
    \vspace{-5em}
    \caption{Here is the result for $\langle A_i(\bm{p}_1)A_j(\bm{p}_2)\rangle$ while setting $i=j$. They are plotted with respect to $|\bm{p}_1-\bm{p}_2|$, for labeling the distance between plane wave bases. The average is taken over an energy shell consisting $N_{ave}$ energy levels.}
    \label{fig:AA}
\end{figure}

Figs.~\ref{fig:AA} and \ref{fig:AAE} show while approaching the semiclassical limit by averaging more energy eigenstates in a wider energy shell, $\langle A_i(\bm{p}_1)A_j(\bm{p}_2)\rangle$ is approaching the $\delta$ function. This means Berry's conjecture predicts the energetic long-wavelength behavior in the space of eigenstates, which means Eq.~\eqref{qx_sq} is the asymptotical behavior of $f$ function while averaging over more eigenstates in the limit $\hbar\to 0$.

\section{Conclusion and Discussion}

Our findings imply that ETH is not only applicable to quantum many-body systems, but also to certain observables in few-body or even single-particle quantum systems such as the quantum billiard, with much fewer degrees of freedom. This implies that ETH is essentially a feature of two point correlation functions of the amplitudes in integrable bases, applying to many quantum chaotic systems, rather than a phenomenon unique to quantum many-body physics, as has often been assumed \cite{deutsch1991quantum, srednicki1994chaos, RevModPhys.91.021001}.

By employing the semiclassical expression of the averaged Wigner function to calculate the $f$ function in ETH, we obtained an analytical expression, Eq.~\eqref{qx_sq}. We demonstrate that this expression is equivalent to the prediction of Berry's conjecture, where the correlation among and within the eigenstates is neglected. Using this analytical expression, we obtain the scaling feature of $f$ function bandwidth and further demonstrate its connection to classical dynamics. This connection suggests that ETH has important physical implications in single-particle or few-body systems that were previously overlooked. Single-particle and few-body models have fewer degrees of freedom, and many of them have well defined classical correspondence, making these models well suited for semiclassical analytical studies of ETH and for exploring the connection between ETH and quantum chaos. Eqs.~\eqref{qx_sq} and \eqref{qx_sq_analy} provide the first explicit analytical solutions for the off-diagonal $f$ function of an observable since the proposal of ETH. The semiclassical method can be extended to many other quantum chaotic systems.

Additionally, by providing numerical evidence, we show that Berry’s conjecture represents the energetic long-wavelength behavior in the space of eigenstates, meaning our analytical expression Eq.~\eqref{qx_sq} is the asymptotical behavior of $f$ function in semiclassical limit. Analytical prediction and numerical simulation for diagonal elements are also provided.

\section{Acknowledgements}
We thank Dr. Chenguang Lyuyin and Dr. Zhenqi Chen for useful guidance about the scaling algorithm, and Prof. Wen-ge Wang for insightful discussions and
helpful feedback on the manuscript. This work is partially supported by the 2024 National Undergraduate Innovation Training Program at University of Science and Technology of China No.202410358024 and the Natural Science Foundation of China under
Grant Nos. 12175222, 11535011, and 11775210.

\appendix
\section{Scaling Relation in Eq.~\eqref{E_i}}
\label{scaling}

In this section, we provide a justification for the scaling relation given in Eq.~\eqref{E_i}.

The time-independent Schrödinger equation for our model reads
\begin{equation}
\nabla^2_{\bm{x}} \psi(\bm{x}) = -k^2 \psi(\bm{x}),
\label{eigen_eq}
\end{equation}
where the energy is related to the wave number via $E = \frac{\hbar^2 k^2}{2m}$.  

If we scale the linear dimensions of the quarter-stadium by a factor $a$, i.e., $\bm{x} \rightarrow \bm{x}' = a \bm{x}$, then Eq.~\eqref{eigen_eq} transforms into
\begin{equation}
\nabla^2_{\bm{x}'} \psi(\bm{x}') = -\frac{k^2}{a^2} \psi(\bm{x}').
\end{equation}
Consequently, the energy scales as
\begin{equation}
E = \frac{\hbar^2 k^2}{2 m a^2}.
\end{equation}

Since the area of the quarter-stadium scales as $S_\Omega \propto a^2$, we find that, for any given energy level $E$, the scaling relation
\begin{equation}
E \propto \frac{\hbar^2}{m S_\Omega}
\end{equation}
holds~\cite{vergini1995calculation}, which explains the relation stated in Eq.~\eqref{E_i}.

\section{Derivation of $S(E)$ in the Quarter-Stadium Billiard}
\label{derivation of S}

In this section, we derive the value for $S(E)$ given in Eq.~\eqref{S} of the main text in our model. The phase space volume for the energy shell is
\begin{equation}
\begin{aligned}
S(E_i) &= \int d\bm{p} \, d\bm{q} \, \delta\Big(\frac{\bm{p}^2}{2m} - E_i\Big)\\
&= \int d\bm{p} \, d\bm{q} \, \sqrt{\frac{m}{2E_i}} \, \delta(|\bm{p}| - \sqrt{2 m E_i})\\
&= 2 \pi m S_\Omega,
\end{aligned}
\end{equation}
where in the last step, the integration domain for $\bm{q}$ is restricted to the quarter-stadium region $\Omega$, whose area is $S_\Omega$, due to the infinite potential walls. In this two-dimensional case, $S(E)$ is constant, so we simply denote $S(E) \equiv S$ in the following.

\section{Derivation of Eq.~\eqref{qx_sq}}
\label{derivation of Eq9}

Here we derive Eq.~\eqref{qx_sq} in the main text. Using Eqs.~\eqref{Osq} and \eqref{J}, we substitute the Wely symbol of the operator and the Wigner function specific to our model:
\begin{align}
&O_{\mathrm{\scriptscriptstyle W}}(\bm{p},\bm{q}) = q_x,\\
&W_i(\bm{p},\bm{q}) = \frac{\delta(\frac{\bm{p}^2}{2m}-E_i)}{S(E_i)},
\end{align}
into Eq.~\eqref{Osq}, yielding
\begin{widetext}
\begin{equation}
\begin{aligned}
|\langle E_i | \hat{q}_x | E_j \rangle|^2 &= \frac{1}{(2\pi\hbar)^2 S^2} \int d\bm{p} d\bm{q} d\bm{p}' d\bm{q}' d\bm{\tilde{p}} d\bm{\tilde{q}} \, 
e^{-\frac{i}{\hbar}\bm{q}'\cdot \bm{\tilde{p}}} e^{-\frac{i}{\hbar}\bm{p}' \cdot \bm{\tilde{q}}} 
\left(q_x^2 - \frac{1}{4} \tilde{q}_x^2\right) \delta\Big(\frac{(\bm{p}-\frac{\bm{p}'}{2})^2}{2m} - E_i\Big) 
\delta\Big(\frac{(\bm{p}+\frac{\bm{p}'}{2})^2}{2m} - E_j\Big)
\end{aligned}
\end{equation}
\end{widetext}

We introduce the coordinate transformation
\begin{equation}
\begin{split}
\bm{\hat{p}} &= \bm{p} - \frac{1}{2}\bm{p}',\\
\bm{\hat{p}'} &= \bm{p} + \frac{1}{2}\bm{p}',
\end{split}
\label{coordinate transformation}
\end{equation}
which maps the integration domain in real space onto the billiard region. It is straightforward to verify that the Jacobian determinant satisfies
\begin{equation}
\det \frac{\partial(\bm{p}, \bm{p'})}{\partial(\bm{\hat{p}}, \bm{\hat{p}'})} = 1.
\end{equation}

Applying this transformation, the matrix element becomes

\begin{equation}
\begin{aligned}
&|\langle E_i | \hat{q}_x | E_j \rangle|^2 = \frac{1}{(2\pi\hbar)^2 S^2} 
\int d\bm{q} d\bm{q}' d\bm{\hat{p}} d\bm{\hat{p}}' d\bm{\tilde{p}} d\bm{\tilde{q}} \,\\
&\quad\times e^{-\frac{i}{\hbar}\bm{q}'\cdot \bm{\tilde{p}}} e^{-\frac{i}{\hbar}(\bm{\hat{p}}' - \bm{\hat{p}}) \cdot \bm{\tilde{q}}} 
\left(q_x^2 - \frac{1}{4} \tilde{q}_x^2\right) \\
&\quad \times \delta\Big(\frac{\bm{\hat{p}}^2}{2m} - E_i\Big) 
\delta\Big(\frac{\bm{\hat{p}}'^2}{2m} - E_j\Big)
\end{aligned}
\end{equation}

Considering the infinite potential walls, the integration domain for $\bm{q}$ and $\bm{q}'$ is restricted to $\bm{q} \pm \frac{\bm{q}'}{2} \in \Omega$. After straightforward algebra, one obtains

\begin{widetext}
\begin{equation}
\begin{aligned}
|\bra{E_i}\hat{q}_x\ket{E_j}|^2&=\frac{1}{4\pi^2S_\Omega^2}\int d\bm{q}d\bm{\tilde{q}}\,(q_x^2-\frac{1}{4}\tilde{q}_x^2)\int d\hat{\theta}\,\exp{\frac{i}{\hbar}\sqrt{2mE_i}|\bm{\tilde{q}}|\cos{(\hat{\theta}}-\tilde{\theta})}\int d\hat{\theta}\,\exp{\frac{i}{\hbar}\sqrt{2mE_i}|\bm{\tilde{q}}|\cos{(\hat{\theta}}-\tilde{\theta})}\\
&=\frac{1}{S_\Omega^2}\int d\bm{q}d\bm{\tilde{q}}\,(q_x^2-\frac{1}{4}\tilde{q}_x^2)J_0(\frac{\sqrt{2mE_i}}{\hbar}|\tilde{q}|)J_0(\frac{\sqrt{2mE_j}}{\hbar}|\tilde{q}|)
\end{aligned}
\label{qx_sq_app}
\end{equation}
\end{widetext}
, where in the last equality we have used the identity~\cite{abramowitz1948handbook} 
\begin{equation}
J_0(x) = \frac{1}{2\pi} \int_0^{2\pi} d\theta \, e^{i x \cos\theta},
\label{J_int}
\end{equation}
and $J_0(x)$ is the zeroth-order Bessel function. Now the integration domain for $\bm{q}$ is $\bm{q}\in\Omega$. By applying the same rescaling as in the main text, one obtains Eq.~\eqref{qx_sq}.

\section{Derivation of Eq.~\eqref{qx_sq_analy}}
\label{derivation of Eq10}

In this section, we derive Eq.~\eqref{qx_sq_analy} in the main text. We employ the asymptotic form of the zeroth-order Bessel function~\cite{abramowitz1948handbook},
\begin{equation}
J_0(x) \approx \sqrt{\frac{2}{\pi x}} \cos\Big(x - \frac{\pi}{4}\Big),
\end{equation}
which provides an excellent approximation for $x \gtrsim 1$. In our case, with $\hbar = 0.01$, this approximation is extremely accurate.

Substituting this into Eq.~\eqref{qx_sq_app}, we obtain
\begin{equation}
\begin{aligned}
&|\langle E_i | \hat{q}_x | E_j \rangle|^2 = \frac{\sqrt{2} \hbar}{\pi \sqrt{m} S_\Omega^2 (E_i E_j)^{1/4}} 
\int d\bm{q} \, d|\bm{\tilde{q}}| \, d\tilde{\theta} \, 
\left(q_x^2 - \frac{1}{4} \tilde{q}_x^2 \right)\\
&\times \cos\Big(\frac{\sqrt{2 m E_i}}{\hbar} |\bm{\tilde{q}}| - \frac{\pi}{4}\Big) 
\cos\Big(\frac{\sqrt{2 m E_j}}{\hbar} |\bm{\tilde{q}}| - \frac{\pi}{4}\Big).
\end{aligned}
\label{qx_sq_as_app}
\end{equation}

Carrying out the integration in Eq.~\eqref{qx_sq_as_app} yields the final expression 
\begin{widetext}
\begin{equation}
\begin{split}
&|\bra{E_i}\hat{q}_x\ket{E_j}|^2=\frac{\hbar^2}{mS_\Omega(E_iE_j)^{\frac{1}{4}}}\biggl\{\left[\frac{1}{S_\Omega}\left(\frac{5\pi}{16}+1\right)-\frac{1}{4}Q^2\right]\biggl(\frac{\sin\left[\frac{\sqrt{2m}}{\hbar}(\sqrt{E_i}-\sqrt{E_j})Q\right]}{\sqrt{E_i}-\sqrt{E_j}}\\
&+\frac{\cos\left[\frac{\sqrt{2m}}{\hbar}(\sqrt{E_i}+\sqrt{E_j})Q\right]}{\sqrt{E_i}+\sqrt{E_j}}\biggr)-\frac{1}{2}\frac{\hbar}{\sqrt{2m}}Q\biggl(\frac{\cos\left[\frac{\sqrt{2m}}{\hbar}(\sqrt{E_i}-\sqrt{E_j})Q\right]}{(\sqrt{E_i}-\sqrt{E_j})^2}-\frac{\sin\left[\frac{\sqrt{2m}}{\hbar}(\sqrt{E_i}+\sqrt{E_j})Q\right]}{(\sqrt{E_i}+\sqrt{E_j})^2}\biggr)\\
&+\frac{1}{2}\left(\frac{\hbar}{\sqrt{2m}}\right)^2\biggl(\frac{\sin\left[\frac{\sqrt{2m}}{\hbar}(\sqrt{E_i}-\sqrt{E_j})Q\right]}{(\sqrt{E_i}-\sqrt{E_j})^3} +\frac{\cos\left[\frac{\sqrt{2m}}{\hbar}(\sqrt{E_i}+\sqrt{E_j})Q\right]}{(\sqrt{E_i}+\sqrt{E_j})^3}\biggr)\biggr\}.
\end{split}
\label{qx_sq_analy_app}
\end{equation}
\end{widetext}

In high energy levels, terms divided by powers of $\sqrt{E_i}+\sqrt{E_j}$ are much smaller than those divided by powers of $\sqrt{E_i}-\sqrt{E_j}$. Neglecting those terms will yield Eq. ~\eqref{qx_sq_analy} in the main text.

\section{Derivation of the Diagonal Function}
In this section, we derive the diagonal function for $q_x$, namely Eq.~\eqref{diag} in the main text.

Using the semiclassical expression Eq.~\eqref{O_diag} \cite{wang2024semiclassicalstudydiagonaloffdiagonal} in the main text, we have
\begin{equation}
    \begin{aligned}
        q_x(E_i)&=\frac{1}{S}\int d\bm{q} d{\bm{p}}\,\delta(\frac{\bm{p}^2}{2m}-E_i)q_x\\
        &=\frac{1}{S}\int d\bm{q} d{\bm{p}}\,\sqrt{\frac{m}{2E_i}}\delta(|\bm{p}|-\sqrt{2mE_i})q_x\\
        &=\frac{2\pi m}{S}\int d\bm{q}\,q_x.
    \end{aligned}
\end{equation}
Here the integration domain for $\bm{q}$ is restricted in the quarter stadium billiard region because of the infinite potential wall. Thus after integration for $\bm{q}$, we can get $q_x(E_i)=\frac{2\pi m}{S}(\frac{5}{6}+\frac{\pi}{4})=\frac{10+3\pi}{12+3\pi}\approx0.967$.

\section{From Berry's Conjecture to Averaged Wigner Function}
\label{Berry and Wigner}
In this section, we demonstrate the equivalence between Berry's conjecture and Eq.~\eqref{qx_sq} in the main text.

Substituting Eq.~\eqref{EigenBerry} into the definition of Wigner function
\begin{equation}
    W_i(\bm{p},\bm{q})=\frac{1}{(2\pi\hbar)^2}\int d\bm{\tilde{p}}\,e^{\frac{i}{\hbar}\bm{\tilde{p}}\cdot\bm{q}}\bra{\bm{p}+\frac{\bm{\tilde{p}}}{2}}\ket{E_i}\bra{E_i}\ket{\bm{p}-\frac{\bm{\tilde{p}}}{2}},
\end{equation}
one get
\begin{widetext}
\begin{equation}
    \begin{aligned}
        \overline{W}_i(\bm{p},\bm{q})&=\frac{\mathcal{N}_i^2}{(2\pi\hbar)^2}\int d\bm{\tilde{p}}\,e^{\frac{i}{\hbar}\bm{\tilde{p}}\cdot\bm{q}}\delta\bigg((\bm{p}+\frac{\bm{\tilde{p}}}{2})^2-2mE_i\bigg)\delta\bigg((\bm{p}-\frac{\bm{\tilde{p}}}{2})^2-2mE_i\bigg)\langle \mathcal{A}_i(\bm{p}+\frac{\bm{\tilde{p}}}{2})\mathcal{A}_i(\bm{p}-\frac{\bm{\tilde{p}}}{2})\rangle\\
        &=\frac{\mathcal{N}_i^2}{(2\pi\hbar)^2}\int d\bm{\tilde{p}}\,e^{\frac{i}{\hbar}\bm{\tilde{p}}\cdot\bm{q}}\delta\bigg((\bm{p}+\frac{\bm{\tilde{p}}}{2})^2-2mE_i\bigg)\delta\bigg((\bm{p}-\frac{\bm{\tilde{p}}}{2})^2-2mE_i\bigg)\delta(\bm{\tilde{p}})/\delta(2\bm{p}\cdot\bm{\tilde{p}})\\
        &=\frac{\mathcal{N}_i^2}{(2\pi\hbar)^2}\delta^2(\bm{p}^2-2mE_i)/\delta(0)\\
        &=\mathcal{\widetilde{N}}_i\,\delta(\frac{\bm{p}^2}{2m}-E_i),
    \end{aligned}  
\end{equation}
\end{widetext}
where $\mathcal{\widetilde{N}}_i$ is the new normalization factor. In Sec. ~\ref{section_NS}, we mentioned that by restricting the integration domain for the observable to be within the billiard, the restriction imposed by the infinite potential wall in the Hamiltonian is automatically satisfied. Thus, using the averaged Wigner function derived from Berry's conjecture can leads to the same result as Eq.~\eqref{qx_sq}.

\section{Divergence of Wigner function in the second-order normalization}
\label{2nd order normalization}
The normalization condition $\braket{E_i|E_j} = \delta_{ij}$, where $\delta_{ij}$ denotes the Kronecker delta, is ensured by introducing the normalization factor $S$ in Eq.~\ref{Wsc}. In this section, however, we derive the relation $|\braket{E_i|E_j}|^2 \propto \delta(E_i - E_j)$, where $\delta(E_i - E_j)$ is the Dirac delta function. Here we refer to the condition $|\braket{E_i|E_j}|^2 = \delta_{ij}$ as the second-order normalization. This result indicates that the second-order normalization of Eq.~\ref{Wsc} breaks down due to the divergence at $E_i = E_j$.

Using the exact relation
\begin{equation}
    \langle \bm{q}_1 | E_i \rangle \langle E_i | \bm{q}_2 \rangle= \int d\bm{p} \, W_i \!\left( \bm{p}, \frac{\bm{q}_1 + \bm{q}_2}{2} \right)e^{- i \bm{p} \cdot (\bm{q}_2 - \bm{q}_1) / \hbar},
\end{equation}
one can get
\begin{equation}
    \begin{aligned}
        &|\bra{E_i}\ket{E_j}|^2=\int d\bm{q}_1d\bm{q}_2\,\bra{\bm{q}_2}\ket{E_i}\bra{E_i}\ket{\bm{q}_1}\bra{\bm{q}_1}\ket{E_j}\bra{E_j}\ket{\bm{q}_2}\\
        &=\int d\bm{q}_1d\bm{q}_2d\bm{p}_1d\bm{p}_2\,W_i(\bm{p}_1,\frac{\bm{q}_1+\bm{q}_2}{2})W_j(\bm{p}_2,\frac{\bm{q}_1+\bm{q}_2}{2})\\
        &\times e^{-\frac{i}{\hbar}(\bm{p}_1-\bm{p}_2)(\bm{q}_1-\bm{q}_2)}.
    \end{aligned}
\end{equation}

After using Eq.~\ref{Wsc} and the following coordinate transformation
\begin{equation}
\begin{split}
\tilde{\bm{q}}_1 &= \frac{\bm{q}_1+\bm{q}_2}{2},\\
\tilde{\bm{q}}_2 &= \bm{q}_1-\bm{q}_2,
\end{split}
\label{coordinate transformation for q}
\end{equation}
one can get (The integration domain for $\tilde{\bm{q}}_1$ will be restricted to the billiard region due to the infinite potential wall.)
\begin{widetext}
\begin{equation}
\begin{aligned}
    |\bra{E_i}\ket{E_j}|^2&=\frac{1}{S^2}\int d\tilde{\bm{q}}_1d\tilde{\bm{q}}_2d\bm{p}_1d\bm{p}_2\,\delta(\frac{\bm{p}_1^2}{2m}-E_i)\delta(\frac{\bm{p}_2^2}{2m}-E_j)e^{-\frac{i}{\hbar}(\bm{p}_1-\bm{p}_2)\cdot\tilde{\bm{q}}_2}\\
    &=\frac{\hbar^2}{m^2S_{\Omega}}\int d\bm{p}_1d\bm{p}_2\,\delta(\frac{\bm{p}_1^2}{2m}-E_i)\delta(\frac{\bm{p}_2^2}{2m}-E_j)\delta(\bm{p}_1-\bm{p}_2)\\
    &=\frac{\hbar^2}{m^2S_{\Omega}}\int d\bm{p}_1\,\delta(\frac{\bm{p}_1^2}{2m}-E_i)\delta(\frac{\bm{p}_1^2}{2m}-E_j)\\
    &=\frac{\hbar^2}{2mS_{\Omega}\sqrt{E_iE_j}}\int d|\bm{p}_1|d\theta_1\,|\bm{p}_1|\delta(|\bm{p}_1|-\sqrt{2mE_i})\delta(|\bm{p}_1|-\sqrt{2mE_j})\\
    &=\frac{\sqrt{2}\pi\hbar^2}{S_{\Omega}\sqrt{mE_i}}\delta\bigg(\frac{\sqrt{2m}}{\sqrt{E_i}+\sqrt{E_j}}(E_i-E_j)\bigg)\\
    &=\frac{2\pi\hbar^2}{mS_\Omega}\delta(E_i-E_j).
\end{aligned} 
\end{equation}
\end{widetext}
\bibliography{ref}

\begin{thebibliography}{44}%
\makeatletter
\providecommand \@ifxundefined [1]{%
 \@ifx{#1\undefined}
}%
\providecommand \@ifnum [1]{%
 \ifnum #1\expandafter \@firstoftwo
 \else \expandafter \@secondoftwo
 \fi
}%
\providecommand \@ifx [1]{%
 \ifx #1\expandafter \@firstoftwo
 \else \expandafter \@secondoftwo
 \fi
}%
\providecommand \natexlab [1]{#1}%
\providecommand \enquote  [1]{``#1''}%
\providecommand \bibnamefont  [1]{#1}%
\providecommand \bibfnamefont [1]{#1}%
\providecommand \citenamefont [1]{#1}%
\providecommand \href@noop [0]{\@secondoftwo}%
\providecommand \href [0]{\begingroup \@sanitize@url \@href}%
\providecommand \@href[1]{\@@startlink{#1}\@@href}%
\providecommand \@@href[1]{\endgroup#1\@@endlink}%
\providecommand \@sanitize@url [0]{\catcode `\\12\catcode `\$12\catcode `\&12\catcode `\#12\catcode `\^12\catcode `\_12\catcode `\%12\relax}%
\providecommand \@@startlink[1]{}%
\providecommand \@@endlink[0]{}%
\providecommand \url  [0]{\begingroup\@sanitize@url \@url }%
\providecommand \@url [1]{\endgroup\@href {#1}{\urlprefix }}%
\providecommand \urlprefix  [0]{URL }%
\providecommand \Eprint [0]{\href }%
\providecommand \doibase [0]{https://doi.org/}%
\providecommand \selectlanguage [0]{\@gobble}%
\providecommand \bibinfo  [0]{\@secondoftwo}%
\providecommand \bibfield  [0]{\@secondoftwo}%
\providecommand \translation [1]{[#1]}%
\providecommand \BibitemOpen [0]{}%
\providecommand \bibitemStop [0]{}%
\providecommand \bibitemNoStop [0]{.\EOS\space}%
\providecommand \EOS [0]{\spacefactor3000\relax}%
\providecommand \BibitemShut  [1]{\csname bibitem#1\endcsname}%
\let\auto@bib@innerbib\@empty
\bibitem [{\citenamefont {Deutsch}(1991)}]{deutsch1991quantum}%
  \BibitemOpen
  \bibfield  {author} {\bibinfo {author} {\bibfnamefont {J.~M.}\ \bibnamefont {Deutsch}},\ }\bibfield  {title} {\bibinfo {title} {Quantum statistical mechanics in a closed system},\ }\href@noop {} {\bibfield  {journal} {\bibinfo  {journal} {Physical review a}\ }\textbf {\bibinfo {volume} {43}},\ \bibinfo {pages} {2046} (\bibinfo {year} {1991})}\BibitemShut {NoStop}%
\bibitem [{\citenamefont {Srednicki}(1994)}]{srednicki1994chaos}%
  \BibitemOpen
  \bibfield  {author} {\bibinfo {author} {\bibfnamefont {M.}~\bibnamefont {Srednicki}},\ }\bibfield  {title} {\bibinfo {title} {Chaos and quantum thermalization},\ }\href@noop {} {\bibfield  {journal} {\bibinfo  {journal} {Physical review e}\ }\textbf {\bibinfo {volume} {50}},\ \bibinfo {pages} {888} (\bibinfo {year} {1994})}\BibitemShut {NoStop}%
\bibitem [{\citenamefont {Rigol}\ \emph {et~al.}(2008)\citenamefont {Rigol}, \citenamefont {Dunjko},\ and\ \citenamefont {Olshanii}}]{rigol2008thermalization}%
  \BibitemOpen
  \bibfield  {author} {\bibinfo {author} {\bibfnamefont {M.}~\bibnamefont {Rigol}}, \bibinfo {author} {\bibfnamefont {V.}~\bibnamefont {Dunjko}},\ and\ \bibinfo {author} {\bibfnamefont {M.}~\bibnamefont {Olshanii}},\ }\bibfield  {title} {\bibinfo {title} {Thermalization and its mechanism for generic isolated quantum systems},\ }\href@noop {} {\bibfield  {journal} {\bibinfo  {journal} {Nature}\ }\textbf {\bibinfo {volume} {452}},\ \bibinfo {pages} {854} (\bibinfo {year} {2008})}\BibitemShut {NoStop}%
\bibitem [{\citenamefont {Beugeling}\ \emph {et~al.}(2014)\citenamefont {Beugeling}, \citenamefont {Moessner},\ and\ \citenamefont {Haque}}]{beugeling2014finite}%
  \BibitemOpen
  \bibfield  {author} {\bibinfo {author} {\bibfnamefont {W.}~\bibnamefont {Beugeling}}, \bibinfo {author} {\bibfnamefont {R.}~\bibnamefont {Moessner}},\ and\ \bibinfo {author} {\bibfnamefont {M.}~\bibnamefont {Haque}},\ }\bibfield  {title} {\bibinfo {title} {Finite-size scaling of eigenstate thermalization},\ }\href@noop {} {\bibfield  {journal} {\bibinfo  {journal} {Physical Review E}\ }\textbf {\bibinfo {volume} {89}},\ \bibinfo {pages} {042112} (\bibinfo {year} {2014})}\BibitemShut {NoStop}%
\bibitem [{\citenamefont {D'Alessio}\ \emph {et~al.}(2016)\citenamefont {D'Alessio}, \citenamefont {Kafri}, \citenamefont {Polkovnikov},\ and\ \citenamefont {Rigol}}]{d2016quantum}%
  \BibitemOpen
  \bibfield  {author} {\bibinfo {author} {\bibfnamefont {L.}~\bibnamefont {D'Alessio}}, \bibinfo {author} {\bibfnamefont {Y.}~\bibnamefont {Kafri}}, \bibinfo {author} {\bibfnamefont {A.}~\bibnamefont {Polkovnikov}},\ and\ \bibinfo {author} {\bibfnamefont {M.}~\bibnamefont {Rigol}},\ }\bibfield  {title} {\bibinfo {title} {From quantum chaos and eigenstate thermalization to statistical mechanics and thermodynamics},\ }\href@noop {} {\bibfield  {journal} {\bibinfo  {journal} {Advances in Physics}\ }\textbf {\bibinfo {volume} {65}},\ \bibinfo {pages} {239} (\bibinfo {year} {2016})}\BibitemShut {NoStop}%
\bibitem [{\citenamefont {Deutsch}(2018)}]{deutsch2018eigenstate}%
  \BibitemOpen
  \bibfield  {author} {\bibinfo {author} {\bibfnamefont {J.~M.}\ \bibnamefont {Deutsch}},\ }\bibfield  {title} {\bibinfo {title} {Eigenstate thermalization hypothesis},\ }\href@noop {} {\bibfield  {journal} {\bibinfo  {journal} {Reports on Progress in Physics}\ }\textbf {\bibinfo {volume} {81}},\ \bibinfo {pages} {082001} (\bibinfo {year} {2018})}\BibitemShut {NoStop}%
\bibitem [{\citenamefont {Mori}\ \emph {et~al.}(2018)\citenamefont {Mori}, \citenamefont {Ikeda}, \citenamefont {Kaminishi},\ and\ \citenamefont {Ueda}}]{Mori_2018}%
  \BibitemOpen
  \bibfield  {author} {\bibinfo {author} {\bibfnamefont {T.}~\bibnamefont {Mori}}, \bibinfo {author} {\bibfnamefont {T.~N.}\ \bibnamefont {Ikeda}}, \bibinfo {author} {\bibfnamefont {E.}~\bibnamefont {Kaminishi}},\ and\ \bibinfo {author} {\bibfnamefont {M.}~\bibnamefont {Ueda}},\ }\bibfield  {title} {\bibinfo {title} {Thermalization and prethermalization in isolated quantum systems: a theoretical overview},\ }\href {https://doi.org/10.1088/1361-6455/aabcdf} {\bibfield  {journal} {\bibinfo  {journal} {Journal of Physics B: Atomic, Molecular and Optical Physics}\ }\textbf {\bibinfo {volume} {51}},\ \bibinfo {pages} {112001} (\bibinfo {year} {2018})}\BibitemShut {NoStop}%
\bibitem [{\citenamefont {Abanin}\ \emph {et~al.}(2019)\citenamefont {Abanin}, \citenamefont {Altman}, \citenamefont {Bloch},\ and\ \citenamefont {Serbyn}}]{RevModPhys.91.021001}%
  \BibitemOpen
  \bibfield  {author} {\bibinfo {author} {\bibfnamefont {D.~A.}\ \bibnamefont {Abanin}}, \bibinfo {author} {\bibfnamefont {E.}~\bibnamefont {Altman}}, \bibinfo {author} {\bibfnamefont {I.}~\bibnamefont {Bloch}},\ and\ \bibinfo {author} {\bibfnamefont {M.}~\bibnamefont {Serbyn}},\ }\bibfield  {title} {\bibinfo {title} {Colloquium: Many-body localization, thermalization, and entanglement},\ }\href {https://doi.org/10.1103/RevModPhys.91.021001} {\bibfield  {journal} {\bibinfo  {journal} {Rev. Mod. Phys.}\ }\textbf {\bibinfo {volume} {91}},\ \bibinfo {pages} {021001} (\bibinfo {year} {2019})}\BibitemShut {NoStop}%
\bibitem [{\citenamefont {Murthy}\ \emph {et~al.}(2023)\citenamefont {Murthy}, \citenamefont {Babakhani}, \citenamefont {Iniguez}, \citenamefont {Srednicki},\ and\ \citenamefont {Yunger~Halpern}}]{PhysRevLett.130.140402}%
  \BibitemOpen
  \bibfield  {author} {\bibinfo {author} {\bibfnamefont {C.}~\bibnamefont {Murthy}}, \bibinfo {author} {\bibfnamefont {A.}~\bibnamefont {Babakhani}}, \bibinfo {author} {\bibfnamefont {F.}~\bibnamefont {Iniguez}}, \bibinfo {author} {\bibfnamefont {M.}~\bibnamefont {Srednicki}},\ and\ \bibinfo {author} {\bibfnamefont {N.}~\bibnamefont {Yunger~Halpern}},\ }\bibfield  {title} {\bibinfo {title} {Non-abelian eigenstate thermalization hypothesis},\ }\href {https://doi.org/10.1103/PhysRevLett.130.140402} {\bibfield  {journal} {\bibinfo  {journal} {Phys. Rev. Lett.}\ }\textbf {\bibinfo {volume} {130}},\ \bibinfo {pages} {140402} (\bibinfo {year} {2023})}\BibitemShut {NoStop}%
\bibitem [{\citenamefont {Pappalardi}\ \emph {et~al.}(2025)\citenamefont {Pappalardi}, \citenamefont {Fritzsch},\ and\ \citenamefont {Prosen}}]{PhysRevLett.134.140404}%
  \BibitemOpen
  \bibfield  {author} {\bibinfo {author} {\bibfnamefont {S.}~\bibnamefont {Pappalardi}}, \bibinfo {author} {\bibfnamefont {F.}~\bibnamefont {Fritzsch}},\ and\ \bibinfo {author} {\bibfnamefont {T.~c.~v.}\ \bibnamefont {Prosen}},\ }\bibfield  {title} {\bibinfo {title} {Full eigenstate thermalization via free cumulants in quantum lattice systems},\ }\href {https://doi.org/10.1103/PhysRevLett.134.140404} {\bibfield  {journal} {\bibinfo  {journal} {Phys. Rev. Lett.}\ }\textbf {\bibinfo {volume} {134}},\ \bibinfo {pages} {140404} (\bibinfo {year} {2025})}\BibitemShut {NoStop}%
\bibitem [{\citenamefont {Bao}\ and\ \citenamefont {Cheng}(2019)}]{bao2019eigenstate}%
  \BibitemOpen
  \bibfield  {author} {\bibinfo {author} {\bibfnamefont {N.}~\bibnamefont {Bao}}\ and\ \bibinfo {author} {\bibfnamefont {N.}~\bibnamefont {Cheng}},\ }\bibfield  {title} {\bibinfo {title} {Eigenstate thermalization hypothesis and approximate quantum error correction},\ }\href@noop {} {\bibfield  {journal} {\bibinfo  {journal} {Journal of High Energy Physics}\ }\textbf {\bibinfo {volume} {2019}},\ \bibinfo {pages} {1} (\bibinfo {year} {2019})}\BibitemShut {NoStop}%
\bibitem [{\citenamefont {Srednicki}(1999)}]{srednicki1999approach}%
  \BibitemOpen
  \bibfield  {author} {\bibinfo {author} {\bibfnamefont {M.}~\bibnamefont {Srednicki}},\ }\bibfield  {title} {\bibinfo {title} {The approach to thermal equilibrium in quantized chaotic systems},\ }\href@noop {} {\bibfield  {journal} {\bibinfo  {journal} {Journal of Physics A: Mathematical and General}\ }\textbf {\bibinfo {volume} {32}},\ \bibinfo {pages} {1163} (\bibinfo {year} {1999})}\BibitemShut {NoStop}%
\bibitem [{\citenamefont {Kaufman}\ \emph {et~al.}(2016)\citenamefont {Kaufman}, \citenamefont {Tai}, \citenamefont {Lukin}, \citenamefont {Rispoli}, \citenamefont {Schittko}, \citenamefont {Preiss},\ and\ \citenamefont {Greiner}}]{kaufman2016quantum}%
  \BibitemOpen
  \bibfield  {author} {\bibinfo {author} {\bibfnamefont {A.~M.}\ \bibnamefont {Kaufman}}, \bibinfo {author} {\bibfnamefont {M.~E.}\ \bibnamefont {Tai}}, \bibinfo {author} {\bibfnamefont {A.}~\bibnamefont {Lukin}}, \bibinfo {author} {\bibfnamefont {M.}~\bibnamefont {Rispoli}}, \bibinfo {author} {\bibfnamefont {R.}~\bibnamefont {Schittko}}, \bibinfo {author} {\bibfnamefont {P.~M.}\ \bibnamefont {Preiss}},\ and\ \bibinfo {author} {\bibfnamefont {M.}~\bibnamefont {Greiner}},\ }\bibfield  {title} {\bibinfo {title} {Quantum thermalization through entanglement in an isolated many-body system},\ }\href@noop {} {\bibfield  {journal} {\bibinfo  {journal} {Science}\ }\textbf {\bibinfo {volume} {353}},\ \bibinfo {pages} {794} (\bibinfo {year} {2016})}\BibitemShut {NoStop}%
\bibitem [{\citenamefont {Clos}\ \emph {et~al.}(2016)\citenamefont {Clos}, \citenamefont {Porras}, \citenamefont {Warring},\ and\ \citenamefont {Schaetz}}]{clos2016time}%
  \BibitemOpen
  \bibfield  {author} {\bibinfo {author} {\bibfnamefont {G.}~\bibnamefont {Clos}}, \bibinfo {author} {\bibfnamefont {D.}~\bibnamefont {Porras}}, \bibinfo {author} {\bibfnamefont {U.}~\bibnamefont {Warring}},\ and\ \bibinfo {author} {\bibfnamefont {T.}~\bibnamefont {Schaetz}},\ }\bibfield  {title} {\bibinfo {title} {Time-resolved observation of thermalization in an isolated quantum system},\ }\href@noop {} {\bibfield  {journal} {\bibinfo  {journal} {Physical review letters}\ }\textbf {\bibinfo {volume} {117}},\ \bibinfo {pages} {170401} (\bibinfo {year} {2016})}\BibitemShut {NoStop}%
\bibitem [{\citenamefont {Turner}\ \emph {et~al.}(2018)\citenamefont {Turner}, \citenamefont {Michailidis}, \citenamefont {Abanin}, \citenamefont {Serbyn},\ and\ \citenamefont {Papi{\'c}}}]{turner2018weak}%
  \BibitemOpen
  \bibfield  {author} {\bibinfo {author} {\bibfnamefont {C.~J.}\ \bibnamefont {Turner}}, \bibinfo {author} {\bibfnamefont {A.~A.}\ \bibnamefont {Michailidis}}, \bibinfo {author} {\bibfnamefont {D.~A.}\ \bibnamefont {Abanin}}, \bibinfo {author} {\bibfnamefont {M.}~\bibnamefont {Serbyn}},\ and\ \bibinfo {author} {\bibfnamefont {Z.}~\bibnamefont {Papi{\'c}}},\ }\bibfield  {title} {\bibinfo {title} {Weak ergodicity breaking from quantum many-body scars},\ }\href@noop {} {\bibfield  {journal} {\bibinfo  {journal} {Nature Physics}\ }\textbf {\bibinfo {volume} {14}},\ \bibinfo {pages} {745} (\bibinfo {year} {2018})}\BibitemShut {NoStop}%
\bibitem [{\citenamefont {Jansen}\ \emph {et~al.}(2019)\citenamefont {Jansen}, \citenamefont {Stolpp}, \citenamefont {Vidmar},\ and\ \citenamefont {Heidrich-Meisner}}]{jansen2019eigenstate}%
  \BibitemOpen
  \bibfield  {author} {\bibinfo {author} {\bibfnamefont {D.}~\bibnamefont {Jansen}}, \bibinfo {author} {\bibfnamefont {J.}~\bibnamefont {Stolpp}}, \bibinfo {author} {\bibfnamefont {L.}~\bibnamefont {Vidmar}},\ and\ \bibinfo {author} {\bibfnamefont {F.}~\bibnamefont {Heidrich-Meisner}},\ }\bibfield  {title} {\bibinfo {title} {Eigenstate thermalization and quantum chaos in the holstein polaron model},\ }\href@noop {} {\bibfield  {journal} {\bibinfo  {journal} {Physical Review B}\ }\textbf {\bibinfo {volume} {99}},\ \bibinfo {pages} {155130} (\bibinfo {year} {2019})}\BibitemShut {NoStop}%
\bibitem [{\citenamefont {LeBlond}\ \emph {et~al.}(2021)\citenamefont {LeBlond}, \citenamefont {Sels}, \citenamefont {Polkovnikov},\ and\ \citenamefont {Rigol}}]{leblond2021universality}%
  \BibitemOpen
  \bibfield  {author} {\bibinfo {author} {\bibfnamefont {T.}~\bibnamefont {LeBlond}}, \bibinfo {author} {\bibfnamefont {D.}~\bibnamefont {Sels}}, \bibinfo {author} {\bibfnamefont {A.}~\bibnamefont {Polkovnikov}},\ and\ \bibinfo {author} {\bibfnamefont {M.}~\bibnamefont {Rigol}},\ }\bibfield  {title} {\bibinfo {title} {Universality in the onset of quantum chaos in many-body systems},\ }\href@noop {} {\bibfield  {journal} {\bibinfo  {journal} {Physical Review B}\ }\textbf {\bibinfo {volume} {104}},\ \bibinfo {pages} {L201117} (\bibinfo {year} {2021})}\BibitemShut {NoStop}%
\bibitem [{\citenamefont {Brenes}\ \emph {et~al.}(2020)\citenamefont {Brenes}, \citenamefont {LeBlond}, \citenamefont {Goold},\ and\ \citenamefont {Rigol}}]{brenes2020eigenstate}%
  \BibitemOpen
  \bibfield  {author} {\bibinfo {author} {\bibfnamefont {M.}~\bibnamefont {Brenes}}, \bibinfo {author} {\bibfnamefont {T.}~\bibnamefont {LeBlond}}, \bibinfo {author} {\bibfnamefont {J.}~\bibnamefont {Goold}},\ and\ \bibinfo {author} {\bibfnamefont {M.}~\bibnamefont {Rigol}},\ }\bibfield  {title} {\bibinfo {title} {Eigenstate thermalization in a locally perturbed integrable system},\ }\href@noop {} {\bibfield  {journal} {\bibinfo  {journal} {Physical review letters}\ }\textbf {\bibinfo {volume} {125}},\ \bibinfo {pages} {070605} (\bibinfo {year} {2020})}\BibitemShut {NoStop}%
\bibitem [{\citenamefont {Mondaini}\ and\ \citenamefont {Rigol}(2017)}]{mondaini2017eigenstate}%
  \BibitemOpen
  \bibfield  {author} {\bibinfo {author} {\bibfnamefont {R.}~\bibnamefont {Mondaini}}\ and\ \bibinfo {author} {\bibfnamefont {M.}~\bibnamefont {Rigol}},\ }\bibfield  {title} {\bibinfo {title} {Eigenstate thermalization in the two-dimensional transverse field ising model. ii. off-diagonal matrix elements of observables},\ }\href@noop {} {\bibfield  {journal} {\bibinfo  {journal} {Physical Review E}\ }\textbf {\bibinfo {volume} {96}},\ \bibinfo {pages} {012157} (\bibinfo {year} {2017})}\BibitemShut {NoStop}%
\bibitem [{\citenamefont {LeBlond}\ \emph {et~al.}(2019)\citenamefont {LeBlond}, \citenamefont {Mallayya}, \citenamefont {Vidmar},\ and\ \citenamefont {Rigol}}]{leblond2019entanglement}%
  \BibitemOpen
  \bibfield  {author} {\bibinfo {author} {\bibfnamefont {T.}~\bibnamefont {LeBlond}}, \bibinfo {author} {\bibfnamefont {K.}~\bibnamefont {Mallayya}}, \bibinfo {author} {\bibfnamefont {L.}~\bibnamefont {Vidmar}},\ and\ \bibinfo {author} {\bibfnamefont {M.}~\bibnamefont {Rigol}},\ }\bibfield  {title} {\bibinfo {title} {Entanglement and matrix elements of observables in interacting integrable systems},\ }\href@noop {} {\bibfield  {journal} {\bibinfo  {journal} {Physical Review E}\ }\textbf {\bibinfo {volume} {100}},\ \bibinfo {pages} {062134} (\bibinfo {year} {2019})}\BibitemShut {NoStop}%
\bibitem [{\citenamefont {Richter}\ \emph {et~al.}(2020)\citenamefont {Richter}, \citenamefont {Dymarsky}, \citenamefont {Steinigeweg},\ and\ \citenamefont {Gemmer}}]{richter2020eigenstate}%
  \BibitemOpen
  \bibfield  {author} {\bibinfo {author} {\bibfnamefont {J.}~\bibnamefont {Richter}}, \bibinfo {author} {\bibfnamefont {A.}~\bibnamefont {Dymarsky}}, \bibinfo {author} {\bibfnamefont {R.}~\bibnamefont {Steinigeweg}},\ and\ \bibinfo {author} {\bibfnamefont {J.}~\bibnamefont {Gemmer}},\ }\bibfield  {title} {\bibinfo {title} {Eigenstate thermalization hypothesis beyond standard indicators: Emergence of random-matrix behavior at small frequencies},\ }\href@noop {} {\bibfield  {journal} {\bibinfo  {journal} {Physical Review E}\ }\textbf {\bibinfo {volume} {102}},\ \bibinfo {pages} {042127} (\bibinfo {year} {2020})}\BibitemShut {NoStop}%
\bibitem [{\citenamefont {Rigol}\ and\ \citenamefont {Srednicki}(2012)}]{rigol2012alternatives}%
  \BibitemOpen
  \bibfield  {author} {\bibinfo {author} {\bibfnamefont {M.}~\bibnamefont {Rigol}}\ and\ \bibinfo {author} {\bibfnamefont {M.}~\bibnamefont {Srednicki}},\ }\bibfield  {title} {\bibinfo {title} {Alternatives to eigenstate thermalization},\ }\href@noop {} {\bibfield  {journal} {\bibinfo  {journal} {Physical review letters}\ }\textbf {\bibinfo {volume} {108}},\ \bibinfo {pages} {110601} (\bibinfo {year} {2012})}\BibitemShut {NoStop}%
\bibitem [{\citenamefont {De~Palma}\ \emph {et~al.}(2015)\citenamefont {De~Palma}, \citenamefont {Serafini}, \citenamefont {Giovannetti},\ and\ \citenamefont {Cramer}}]{de2015necessity}%
  \BibitemOpen
  \bibfield  {author} {\bibinfo {author} {\bibfnamefont {G.}~\bibnamefont {De~Palma}}, \bibinfo {author} {\bibfnamefont {A.}~\bibnamefont {Serafini}}, \bibinfo {author} {\bibfnamefont {V.}~\bibnamefont {Giovannetti}},\ and\ \bibinfo {author} {\bibfnamefont {M.}~\bibnamefont {Cramer}},\ }\bibfield  {title} {\bibinfo {title} {Necessity of eigenstate thermalization},\ }\href@noop {} {\bibfield  {journal} {\bibinfo  {journal} {Physical review letters}\ }\textbf {\bibinfo {volume} {115}},\ \bibinfo {pages} {220401} (\bibinfo {year} {2015})}\BibitemShut {NoStop}%
\bibitem [{\citenamefont {Bernien}\ \emph {et~al.}(2017)\citenamefont {Bernien}, \citenamefont {Schwartz}, \citenamefont {Keesling}, \citenamefont {Levine}, \citenamefont {Omran}, \citenamefont {Pichler}, \citenamefont {Choi}, \citenamefont {Zibrov}, \citenamefont {Endres}, \citenamefont {Greiner} \emph {et~al.}}]{bernien2017probing}%
  \BibitemOpen
  \bibfield  {author} {\bibinfo {author} {\bibfnamefont {H.}~\bibnamefont {Bernien}}, \bibinfo {author} {\bibfnamefont {S.}~\bibnamefont {Schwartz}}, \bibinfo {author} {\bibfnamefont {A.}~\bibnamefont {Keesling}}, \bibinfo {author} {\bibfnamefont {H.}~\bibnamefont {Levine}}, \bibinfo {author} {\bibfnamefont {A.}~\bibnamefont {Omran}}, \bibinfo {author} {\bibfnamefont {H.}~\bibnamefont {Pichler}}, \bibinfo {author} {\bibfnamefont {S.}~\bibnamefont {Choi}}, \bibinfo {author} {\bibfnamefont {A.~S.}\ \bibnamefont {Zibrov}}, \bibinfo {author} {\bibfnamefont {M.}~\bibnamefont {Endres}}, \bibinfo {author} {\bibfnamefont {M.}~\bibnamefont {Greiner}}, \emph {et~al.},\ }\bibfield  {title} {\bibinfo {title} {Probing many-body dynamics on a 51-atom quantum simulator},\ }\href@noop {} {\bibfield  {journal} {\bibinfo  {journal} {Nature}\ }\textbf {\bibinfo {volume} {551}},\ \bibinfo {pages} {579} (\bibinfo {year} {2017})}\BibitemShut {NoStop}%
\bibitem [{\citenamefont {Garrison}\ and\ \citenamefont {Grover}(2018)}]{garrison2018does}%
  \BibitemOpen
  \bibfield  {author} {\bibinfo {author} {\bibfnamefont {J.~R.}\ \bibnamefont {Garrison}}\ and\ \bibinfo {author} {\bibfnamefont {T.}~\bibnamefont {Grover}},\ }\bibfield  {title} {\bibinfo {title} {Does a single eigenstate encode the full hamiltonian?},\ }\href@noop {} {\bibfield  {journal} {\bibinfo  {journal} {Physical Review X}\ }\textbf {\bibinfo {volume} {8}},\ \bibinfo {pages} {021026} (\bibinfo {year} {2018})}\BibitemShut {NoStop}%
\bibitem [{\citenamefont {Khatami}\ \emph {et~al.}(2013)\citenamefont {Khatami}, \citenamefont {Pupillo}, \citenamefont {Srednicki},\ and\ \citenamefont {Rigol}}]{khatami2013fluctuation}%
  \BibitemOpen
  \bibfield  {author} {\bibinfo {author} {\bibfnamefont {E.}~\bibnamefont {Khatami}}, \bibinfo {author} {\bibfnamefont {G.}~\bibnamefont {Pupillo}}, \bibinfo {author} {\bibfnamefont {M.}~\bibnamefont {Srednicki}},\ and\ \bibinfo {author} {\bibfnamefont {M.}~\bibnamefont {Rigol}},\ }\bibfield  {title} {\bibinfo {title} {Fluctuation-dissipation theorem in an isolated system of quantum<? format?> dipolar bosons after a quench},\ }\href@noop {} {\bibfield  {journal} {\bibinfo  {journal} {Physical review letters}\ }\textbf {\bibinfo {volume} {111}},\ \bibinfo {pages} {050403} (\bibinfo {year} {2013})}\BibitemShut {NoStop}%
\bibitem [{\citenamefont {Steinigeweg}\ \emph {et~al.}(2013)\citenamefont {Steinigeweg}, \citenamefont {Herbrych},\ and\ \citenamefont {Prelov{\v{s}}ek}}]{steinigeweg2013eigenstate}%
  \BibitemOpen
  \bibfield  {author} {\bibinfo {author} {\bibfnamefont {R.}~\bibnamefont {Steinigeweg}}, \bibinfo {author} {\bibfnamefont {J.}~\bibnamefont {Herbrych}},\ and\ \bibinfo {author} {\bibfnamefont {P.}~\bibnamefont {Prelov{\v{s}}ek}},\ }\bibfield  {title} {\bibinfo {title} {Eigenstate thermalization within isolated spin-chain systems},\ }\href@noop {} {\bibfield  {journal} {\bibinfo  {journal} {Physical Review E—Statistical, Nonlinear, and Soft Matter Physics}\ }\textbf {\bibinfo {volume} {87}},\ \bibinfo {pages} {012118} (\bibinfo {year} {2013})}\BibitemShut {NoStop}%
\bibitem [{\citenamefont {Mondaini}\ \emph {et~al.}(2016)\citenamefont {Mondaini}, \citenamefont {Fratus}, \citenamefont {Srednicki},\ and\ \citenamefont {Rigol}}]{mondaini2016eigenstate}%
  \BibitemOpen
  \bibfield  {author} {\bibinfo {author} {\bibfnamefont {R.}~\bibnamefont {Mondaini}}, \bibinfo {author} {\bibfnamefont {K.~R.}\ \bibnamefont {Fratus}}, \bibinfo {author} {\bibfnamefont {M.}~\bibnamefont {Srednicki}},\ and\ \bibinfo {author} {\bibfnamefont {M.}~\bibnamefont {Rigol}},\ }\bibfield  {title} {\bibinfo {title} {Eigenstate thermalization in the two-dimensional transverse field ising model},\ }\href@noop {} {\bibfield  {journal} {\bibinfo  {journal} {Physical Review E}\ }\textbf {\bibinfo {volume} {93}},\ \bibinfo {pages} {032104} (\bibinfo {year} {2016})}\BibitemShut {NoStop}%
\bibitem [{\citenamefont {Dymarsky}(2022)}]{dymarsky2022bound}%
  \BibitemOpen
  \bibfield  {author} {\bibinfo {author} {\bibfnamefont {A.}~\bibnamefont {Dymarsky}},\ }\bibfield  {title} {\bibinfo {title} {Bound on eigenstate thermalization from transport},\ }\href@noop {} {\bibfield  {journal} {\bibinfo  {journal} {Physical Review Letters}\ }\textbf {\bibinfo {volume} {128}},\ \bibinfo {pages} {190601} (\bibinfo {year} {2022})}\BibitemShut {NoStop}%
\bibitem [{\citenamefont {Wang}\ and\ \citenamefont {ge~Wang}(2024)}]{wang2024semiclassicalstudydiagonaloffdiagonal}%
  \BibitemOpen
  \bibfield  {author} {\bibinfo {author} {\bibfnamefont {X.}~\bibnamefont {Wang}}\ and\ \bibinfo {author} {\bibfnamefont {W.}~\bibnamefont {ge~Wang}},\ }\href {https://arxiv.org/abs/2210.13183} {\bibinfo {title} {Semiclassical study of diagonal and offdiagonal functions in the eigenstate thermalization hypothesis}} (\bibinfo {year} {2024}),\ \Eprint {https://arxiv.org/abs/2210.13183} {arXiv:2210.13183 [cond-mat.stat-mech]} \BibitemShut {NoStop}%
\bibitem [{\citenamefont {Wang}\ and\ \citenamefont {ge~Wang}(2025)}]{wang2025operatorweylsymbolapproacheigenstatethermalization}%
  \BibitemOpen
  \bibfield  {author} {\bibinfo {author} {\bibfnamefont {X.}~\bibnamefont {Wang}}\ and\ \bibinfo {author} {\bibfnamefont {W.}~\bibnamefont {ge~Wang}},\ }\href {https://arxiv.org/abs/2509.24490} {\bibinfo {title} {An operator-weyl-symbol approach to eigenstate thermalization hypothesis}} (\bibinfo {year} {2025}),\ \Eprint {https://arxiv.org/abs/2509.24490} {arXiv:2509.24490 [quant-ph]} \BibitemShut {NoStop}%
\bibitem [{\citenamefont {Berry}(1977)}]{berry1977regular}%
  \BibitemOpen
  \bibfield  {author} {\bibinfo {author} {\bibfnamefont {M.~V.}\ \bibnamefont {Berry}},\ }\bibfield  {title} {\bibinfo {title} {Regular and irregular semiclassical wavefunctions},\ }\href@noop {} {\bibfield  {journal} {\bibinfo  {journal} {Journal of Physics A: Mathematical and General}\ }\textbf {\bibinfo {volume} {10}},\ \bibinfo {pages} {2083} (\bibinfo {year} {1977})}\BibitemShut {NoStop}%
\bibitem [{\citenamefont {Bohigas}\ \emph {et~al.}(1984)\citenamefont {Bohigas}, \citenamefont {Giannoni},\ and\ \citenamefont {Schmit}}]{bohigas1984characterization}%
  \BibitemOpen
  \bibfield  {author} {\bibinfo {author} {\bibfnamefont {O.}~\bibnamefont {Bohigas}}, \bibinfo {author} {\bibfnamefont {M.-J.}\ \bibnamefont {Giannoni}},\ and\ \bibinfo {author} {\bibfnamefont {C.}~\bibnamefont {Schmit}},\ }\bibfield  {title} {\bibinfo {title} {Characterization of chaotic quantum spectra and universality of level fluctuation laws},\ }\href@noop {} {\bibfield  {journal} {\bibinfo  {journal} {Physical review letters}\ }\textbf {\bibinfo {volume} {52}},\ \bibinfo {pages} {1} (\bibinfo {year} {1984})}\BibitemShut {NoStop}%
\bibitem [{\citenamefont {Berry}(1989)}]{berry1989quantum}%
  \BibitemOpen
  \bibfield  {author} {\bibinfo {author} {\bibfnamefont {M.~V.}\ \bibnamefont {Berry}},\ }\bibfield  {title} {\bibinfo {title} {Quantum scars of classical closed orbits in phase space},\ }\href@noop {} {\bibfield  {journal} {\bibinfo  {journal} {Proceedings of the Royal Society of London. A. Mathematical and Physical Sciences}\ }\textbf {\bibinfo {volume} {423}},\ \bibinfo {pages} {219} (\bibinfo {year} {1989})}\BibitemShut {NoStop}%
\bibitem [{\citenamefont {Weyl}(1927)}]{weyl1927quantenmechanik}%
  \BibitemOpen
  \bibfield  {author} {\bibinfo {author} {\bibfnamefont {H.}~\bibnamefont {Weyl}},\ }\bibfield  {title} {\bibinfo {title} {Quantenmechanik und gruppentheorie},\ }\href@noop {} {\bibfield  {journal} {\bibinfo  {journal} {Zeitschrift f{\"u}r Physik}\ }\textbf {\bibinfo {volume} {46}},\ \bibinfo {pages} {1} (\bibinfo {year} {1927})}\BibitemShut {NoStop}%
\bibitem [{\citenamefont {Voros}(1976)}]{voros1976semi}%
  \BibitemOpen
  \bibfield  {author} {\bibinfo {author} {\bibfnamefont {A.}~\bibnamefont {Voros}},\ }\bibfield  {title} {\bibinfo {title} {Semi-classical approximations},\ }in\ \href@noop {} {\emph {\bibinfo {booktitle} {Annales de l'institut Henri Poincar{\'e}. Section A, Physique Th{\'e}orique}}},\ Vol.~\bibinfo {volume} {24}\ (\bibinfo {year} {1976})\ pp.\ \bibinfo {pages} {31--90}\BibitemShut {NoStop}%
\bibitem [{\citenamefont {Voros}(1977)}]{voros1977asymptotic}%
  \BibitemOpen
  \bibfield  {author} {\bibinfo {author} {\bibfnamefont {A.}~\bibnamefont {Voros}},\ }\bibfield  {title} {\bibinfo {title} {Asymptotic $\hbar$-expansions of stationary quantum states},\ }in\ \href@noop {} {\emph {\bibinfo {booktitle} {Annales de l'institut Henri Poincar{\'e}. Section A, Physique Th{\'e}orique}}},\ Vol.~\bibinfo {volume} {26}\ (\bibinfo {year} {1977})\ pp.\ \bibinfo {pages} {343--403}\BibitemShut {NoStop}%
\bibitem [{\citenamefont {Wang}\ \emph {et~al.}(2022)\citenamefont {Wang}, \citenamefont {Lamann}, \citenamefont {Richter}, \citenamefont {Steinigeweg}, \citenamefont {Dymarsky},\ and\ \citenamefont {Gemmer}}]{wang2022eigenstate}%
  \BibitemOpen
  \bibfield  {author} {\bibinfo {author} {\bibfnamefont {J.}~\bibnamefont {Wang}}, \bibinfo {author} {\bibfnamefont {M.~H.}\ \bibnamefont {Lamann}}, \bibinfo {author} {\bibfnamefont {J.}~\bibnamefont {Richter}}, \bibinfo {author} {\bibfnamefont {R.}~\bibnamefont {Steinigeweg}}, \bibinfo {author} {\bibfnamefont {A.}~\bibnamefont {Dymarsky}},\ and\ \bibinfo {author} {\bibfnamefont {J.}~\bibnamefont {Gemmer}},\ }\bibfield  {title} {\bibinfo {title} {Eigenstate thermalization hypothesis and its deviations from random-matrix theory beyond the thermalization time},\ }\href@noop {} {\bibfield  {journal} {\bibinfo  {journal} {Physical Review Letters}\ }\textbf {\bibinfo {volume} {128}},\ \bibinfo {pages} {180601} (\bibinfo {year} {2022})}\BibitemShut {NoStop}%
\bibitem [{\citenamefont {Schiulaz}\ \emph {et~al.}(2019)\citenamefont {Schiulaz}, \citenamefont {Torres-Herrera},\ and\ \citenamefont {Santos}}]{schiulaz2019thouless}%
  \BibitemOpen
  \bibfield  {author} {\bibinfo {author} {\bibfnamefont {M.}~\bibnamefont {Schiulaz}}, \bibinfo {author} {\bibfnamefont {E.~J.}\ \bibnamefont {Torres-Herrera}},\ and\ \bibinfo {author} {\bibfnamefont {L.~F.}\ \bibnamefont {Santos}},\ }\bibfield  {title} {\bibinfo {title} {Thouless and relaxation time scales in many-body quantum systems},\ }\href@noop {} {\bibfield  {journal} {\bibinfo  {journal} {Physical Review B}\ }\textbf {\bibinfo {volume} {99}},\ \bibinfo {pages} {174313} (\bibinfo {year} {2019})}\BibitemShut {NoStop}%
\bibitem [{\citenamefont {Serbyn}\ \emph {et~al.}(2017)\citenamefont {Serbyn}, \citenamefont {Papi{\'c}},\ and\ \citenamefont {Abanin}}]{serbyn2017thouless}%
  \BibitemOpen
  \bibfield  {author} {\bibinfo {author} {\bibfnamefont {M.}~\bibnamefont {Serbyn}}, \bibinfo {author} {\bibfnamefont {Z.}~\bibnamefont {Papi{\'c}}},\ and\ \bibinfo {author} {\bibfnamefont {D.~A.}\ \bibnamefont {Abanin}},\ }\bibfield  {title} {\bibinfo {title} {Thouless energy and multifractality across the many-body localization transition},\ }\href@noop {} {\bibfield  {journal} {\bibinfo  {journal} {Physical Review B}\ }\textbf {\bibinfo {volume} {96}},\ \bibinfo {pages} {104201} (\bibinfo {year} {2017})}\BibitemShut {NoStop}%
\bibitem [{\citenamefont {{\v{S}}untajs}\ \emph {et~al.}(2020)\citenamefont {{\v{S}}untajs}, \citenamefont {Bon{\v{c}}a}, \citenamefont {Prosen},\ and\ \citenamefont {Vidmar}}]{vsuntajs2020quantum}%
  \BibitemOpen
  \bibfield  {author} {\bibinfo {author} {\bibfnamefont {J.}~\bibnamefont {{\v{S}}untajs}}, \bibinfo {author} {\bibfnamefont {J.}~\bibnamefont {Bon{\v{c}}a}}, \bibinfo {author} {\bibfnamefont {T.}~\bibnamefont {Prosen}},\ and\ \bibinfo {author} {\bibfnamefont {L.}~\bibnamefont {Vidmar}},\ }\bibfield  {title} {\bibinfo {title} {Quantum chaos challenges many-body localization},\ }\href@noop {} {\bibfield  {journal} {\bibinfo  {journal} {Physical Review E}\ }\textbf {\bibinfo {volume} {102}},\ \bibinfo {pages} {062144} (\bibinfo {year} {2020})}\BibitemShut {NoStop}%
\bibitem [{\citenamefont {Kos}\ \emph {et~al.}(2021)\citenamefont {Kos}, \citenamefont {Bertini},\ and\ \citenamefont {Prosen}}]{kos2021chaos}%
  \BibitemOpen
  \bibfield  {author} {\bibinfo {author} {\bibfnamefont {P.}~\bibnamefont {Kos}}, \bibinfo {author} {\bibfnamefont {B.}~\bibnamefont {Bertini}},\ and\ \bibinfo {author} {\bibfnamefont {T.}~\bibnamefont {Prosen}},\ }\bibfield  {title} {\bibinfo {title} {Chaos and ergodicity in extended quantum systems with noisy driving},\ }\href@noop {} {\bibfield  {journal} {\bibinfo  {journal} {Physical review letters}\ }\textbf {\bibinfo {volume} {126}},\ \bibinfo {pages} {190601} (\bibinfo {year} {2021})}\BibitemShut {NoStop}%
\bibitem [{\citenamefont {Vergini}\ and\ \citenamefont {Saraceno}(1995)}]{vergini1995calculation}%
  \BibitemOpen
  \bibfield  {author} {\bibinfo {author} {\bibfnamefont {E.}~\bibnamefont {Vergini}}\ and\ \bibinfo {author} {\bibfnamefont {M.}~\bibnamefont {Saraceno}},\ }\bibfield  {title} {\bibinfo {title} {Calculation by scaling of highly excited states of billiards},\ }\href@noop {} {\bibfield  {journal} {\bibinfo  {journal} {Physical Review E}\ }\textbf {\bibinfo {volume} {52}},\ \bibinfo {pages} {2204} (\bibinfo {year} {1995})}\BibitemShut {NoStop}%
\bibitem [{\citenamefont {Abramowitz}\ and\ \citenamefont {Stegun}(1948)}]{abramowitz1948handbook}%
  \BibitemOpen
  \bibfield  {author} {\bibinfo {author} {\bibfnamefont {M.}~\bibnamefont {Abramowitz}}\ and\ \bibinfo {author} {\bibfnamefont {I.~A.}\ \bibnamefont {Stegun}},\ }\href@noop {} {\emph {\bibinfo {title} {Handbook of mathematical functions with formulas, graphs, and mathematical tables}}},\ Vol.~\bibinfo {volume} {55}\ (\bibinfo  {publisher} {US Government printing office},\ \bibinfo {year} {1948})\BibitemShut {NoStop}%
\end{thebibliography}%
\end{document}